\documentclass[prl,twocolumn,superscriptaddress,floatfix,noeprint]{revtex4-2}

\usepackage[caption=false]{subfig}
\usepackage{graphicx,tabularx,makecell}
\usepackage{amsmath,amsfonts,amssymb}
\usepackage{mathtools}
\usepackage{hhline}
\usepackage{bm,dsfont}
\usepackage[dvipsnames]{xcolor}
\usepackage[colorlinks=true, citecolor=Green, linkcolor=BrickRed, urlcolor=NavyBlue]{hyperref}
\usepackage[all]{hypcap}
\usepackage[utf8]{inputenc}
\usepackage[T1]{fontenc}
\usepackage[normalem]{ulem}

\graphicspath{{figures/}}

\newcommand{\G}{\bm{g}}

\begin{document}

\title{Elastic Screening of Pseudogauge Fields in Graphene}

\author{Christophe De Beule}
\affiliation{Department of Physics and Astronomy, University of Pennsylvania, Philadelphia, Pennsylvania 19104, USA}
\author{Robin Smeyers}
\affiliation{Department of Physics and NANOlight Center of Excellence, University of Antwerp, Groenenborgerlaan 171, 2020 Antwerp, Belgium}
\author{Wilson Nieto Luna}
\affiliation{Department of Physics and NANOlight Center of Excellence, University of Antwerp, Groenenborgerlaan 171, 2020 Antwerp, Belgium}
\author{E. J. Mele}
\affiliation{Department of Physics and Astronomy, University of Pennsylvania, Philadelphia, Pennsylvania 19104, USA}
\author{Lucian Covaci}
\affiliation{Department of Physics and NANOlight Center of Excellence, University of Antwerp, Groenenborgerlaan 171, 2020 Antwerp, Belgium}

\date{\today}

\begin{abstract}
Lattice deformations in graphene couple to the low-energy electronic degrees of freedom as effective scalar and gauge fields. Using molecular dynamics simulations, we show that the optical component of the displacement field, i.e., the relative motion of different sublattices, contributes at equal order as the acoustic component and effectively screens the pseudogauge fields. In particular, we consider twisted bilayer graphene and corrugated monolayer graphene. In both cases, optical lattice displacements significantly reduce the overall magnitude of the pseudomagnetic fields. For corrugated graphene, optical contributions also reshape the pseudomagnetic field and significantly modify the electronic bands near charge neutrality. Previous studies based on continuum elasticity, which ignores this effect, have therefore systematically overestimated the strength of the strain-induced pseudomagnetic field. Our results have important consequences for the interpretation of experiments and design of straintronic applications.
\end{abstract}

\maketitle

It is well known that lattice deformations in graphene couple to the low-energy electronic degrees of freedom as effective scalar and gauge fields \cite{katsnelson_graphene_2007,vozmediano_gauge_2010,amorim_novel_2016}. Intuitively, this can be understood from local symmetry breaking which shifts the Dirac cones near charge neutrality both in energy and momentum, respectively. Indeed, the microscopic $\mathcal C_{3z}$ symmetry of pristine graphene pins the two Dirac points at the zone corners (valleys) of the Brillouin zone \cite{bernevig_chapter_2013}. Atomic displacements that break this symmetry, i.e., shear strain, therefore result in a spatially-varying shift of the Dirac point. This is the action of a vector potential with opposite sign in the two valleys to preserve time-reversal symmetry. For specific strain configurations, the corresponding pseudomagnetic fields give rise to pseudo Landau levels \cite{guinea_generating_2010,guinea_energy_2010,low_gaps_2011} with field strengths that can exceed several hundreds of Tesla \cite{levy_strain-induced_2010,nigge_room_2019}. Moreover, similar pseudogauge fields also arise in strained 2D semiconductors \cite{roldan_strain_2015,cazalilla_quantum_2014,rostami_theory_2015}, as well as in 3D topological semimetals \cite{cortijo_elastic_2015,pikulin_chiral_2016,ilan_pseudo-electromagnetic_2020}. In fact, pseudogauge fields were first considered in semiconductors in the 1980s \cite{iordanskii_dislocations_1985}. In the context of graphene, they were first discussed in carbon nanotubes where the curvature of the tube gives rise to a pseudogauge field that results in a band gap for nanotubes that should otherwise be metallic \cite{kane_size_1997}.

\begin{figure}
    \centering
    \includegraphics[width=\linewidth]{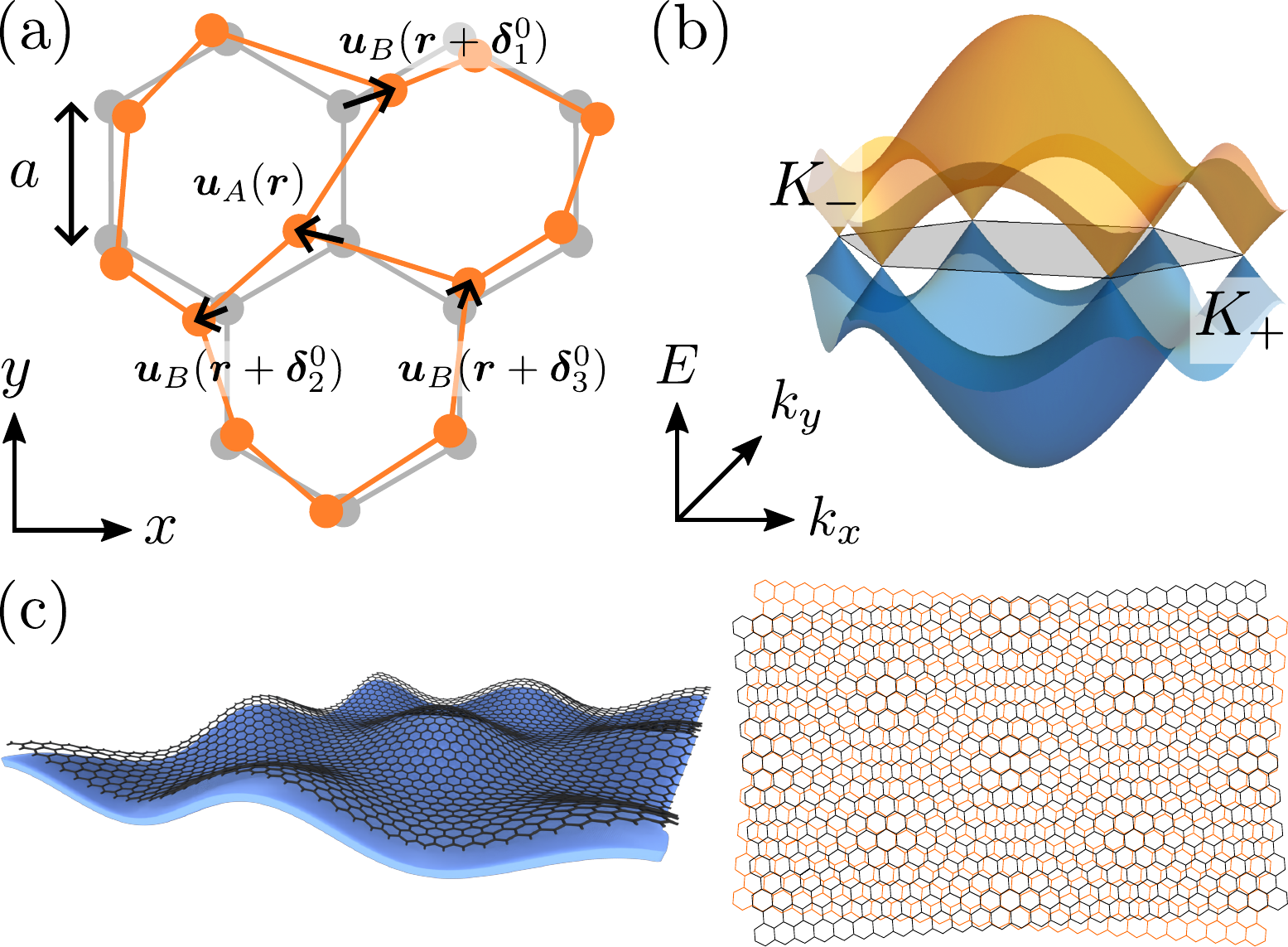}
    \caption{(a) Pristine (gray) and deformed (orange) graphene lattice showing atomic displacements, exaggerated here for clarity. (b) Electronic bands of $p_z$ electrons in graphene around charge neutrality, showing the two Dirac cones at $K_\pm$. (c) Illustration of corrugated monolayer graphene (left) and bilayer moir\'e graphene (right).}
    \label{fig:fig1}
\end{figure}

In most electronic continuum theories, the pseudogauge field is derived from deformations obtained from continuum elasticity which only accounts for acoustic displacements. Since graphene has two sublattices [see Fig.\ \ref{fig:fig1}(a)], there are both center-of-mass (acoustic) and relative (optical) displacements, and both are important in the long-wavelength limit. In this Letter, we derive the contribution to the effective vector potential for a general optical displacement field and show that it contributes at the same order of magnitude as the acoustic shear strains. This finding resolves a long-standing conundrum from earlier works that combine molecular dynamics (MD) simulations and electronic tight-binding models, for which the pseudomagnetic field is found to be much smaller as predicted by elastic theory \cite{neek-amal_nanoengineered_2012,neek-amal_electronic_2013,zhu_programmable_2015,shi_large-area_2020,hsu_nanoscale_2020}.

We demonstrate our theory for two graphene systems: twisted bilayer graphene and corrugated monolayer graphene, see Fig.\ \ref{fig:fig1}(c), for which the displacement fields are computed from MD simulations with the \textsc{lammps} code \cite{thompson_lammps_2022}. In particular, for twisted bilayer graphene near the magic angle we show that optical displacements due to lattice relaxation reduce the pseudomagnetic field by almost one order of magnitude. 

\textcolor{NavyBlue}{\emph{Pseudogauge fields in graphene revisited.}}---Consider a sheet of graphene subject to atomic displacement fields that vary slowly relative to the lattice. These may be induced by external stresses or lattice relaxation, e.g., in a graphene moir\'e. In the long-wavelength limit, we define fields $\bm u_\sigma(\bm r)$ and $h_\sigma(\bm r)$ for sublattice $\sigma = A,B$ [see Fig.\ref{fig:fig1}(a)] projected on and normal to the nominal graphene $xy$ plane, respectively. The long wavelength acoustic and optical displacement fields are defined as
\begin{alignat}{2}
    \bm u & = ( \bm u_A + \bm u_B ) / 2, \qquad && h = ( h_A + h_B ) / 2, \\
    \bm v & = \bm u_A - \bm u_B, \qquad && w = h_A - h_B. \label{eq:op}
\end{alignat}

Atomic displacements modulate the electronic hopping amplitude, which couples to the low-energy Dirac electrons as effective scalar and gauge fields. We revisit this theory, starting from a tight-binding description for the $p_z$ electrons of graphene in the nearest-neighbor approximation. We show that there is an important contribution to the pseudogauge field from optical displacements [$\bm v$ and $w$ in Eq.\ \eqref{eq:op}] that was not considered in previous works and which cannot be obtained from continuum elasticity \cite{amorim_novel_2016}. Here we do not consider the scalar deformation potential \cite{choi_effects_2010,low_gaps_2011,van_der_donck_piezoelectricity_2016,grassano_work_2020}, which to leading order only depends on acoustic displacements as it enters through on-site potentials. Moreover, the deformation potential gives rise to electron-hole puddles and is strongly screened \cite{fogler_pseudomagnetic_2008,wehling_midgap_2008,park_electronphonon_2014,sohier_phonon-limited_2014}.

In the presence of atomic displacements, the Hamiltonian for $p_z$ electrons in graphene can be written as
\begin{equation} \label{eq:tb}
    H = -\sum_{\bm r} \sum_{n=1}^3 t_n(\bm r) c_A^\dag(\bm r) c_B\left( \bm r + \bm \delta_n^0 \right) + \text{h.c.}, 
\end{equation}
where the sums run over cells $\bm r$ and nearest neighbors $n$, and we use the original positions to label atomic sites. Here $c_\sigma^\dag(\bm r)$ [$c_\sigma(\bm r)$] are electron creation (annihilation) operators. The position of $A$ atoms is given by $\bm r + \bm u_A(\bm r) + h_A(\bm r) \hat z$ and $\bm \delta_n(\bm r) = \bm \delta_n^0 + \bm u_B(\bm r + \bm \delta_n^0) - \bm u_A(\bm r) + [ h_B(\bm r + \bm \delta_n^0) - h_A(\bm r) ] \hat z$ are the nearest-neighbor bond vectors with $\bm \delta_n^0$ those of pristine graphene, see Fig.\ \ref{fig:fig1}(a). Taking the continuum limit, one obtains an effective low-energy Hamiltonian \cite{vozmediano_gauge_2010}
\begin{equation} \label{eq:Heff}
    H_\text{eff} = \hbar v_F \sum_\tau \int d^2 \bm r \, \psi_\tau^\dag(\bm r) \left[ -i \nabla + \frac{e\tau}{\hbar} \bm A(\bm r) \right] \cdot \bm \sigma  \psi_\tau(\bm r),
\end{equation}
with $\hbar v_F = \sqrt{3}t_0a/2$, $\bm \sigma = (\tau \sigma_x, \sigma_y)$, and field operators $\psi_\tau(\bm r) = \left[ \psi_{\tau A}(\bm r), \psi_{\tau B}(\bm r) \right]^t$ where $\tau = \pm 1$ is the valley index. See Supplemental Material (SM) for details \footnote{See Supplemental Material [url] for more details on the derivation, \textsc{lammps} simulations, and continuum elasticity for the corrugated graphene. This also includes Refs.\ \cite{haldane_model_1988,nelson_david_statistical_2004,sabio_electrostatic_2008,lee_band_2017,ramires_electrically_2018,lopez-bezanilla_electrical_2020,wang_exact_2021}. For twisted bilayer graphene, we also compare different molecular dynamics potentials \cite{kolmogorov_registry-dependent_2005,brenner_second-generation_2002,rowe_accurate_2020,pallewela_private_nodate}. We also show the optical contribution using old data for hexagonal graphene flakes under triaxial stress \cite{neek-amal_electronic_2013}.}. Here we take the zigzag direction along the $x$ axis. In this case, the effective vector potential is defined as
\begin{equation} \label{eq:Adef}
    A_x(\bm r) - i A_y(\bm r) = -\frac{1}{ev_F} \sum_{n=1}^3 \delta t_n(\bm r) e^{i\bm K \cdot \bm \delta_n^0},
\end{equation}
where $\delta t_n = t_n - t_0$ is the change in hopping with $t_0 \approx 2.8 \; \text{eV}$ and $\bm K_+ = 4\pi/(3a) \hat x$ is the zone corner [see Fig.\ \ref{fig:fig1}(b)] with $a \approx 2.46 \, \text{\r A}$ the graphene lattice constant \cite{castro_neto_electronic_2009}. In lowest order of displacements and their gradients,
\begin{equation} \label{eq:delta}
    \bm \delta_n(\bm r)- \bm \delta_n^0 \simeq \left( \bm \delta_n^0 \cdot \nabla \right) \left( \bm u+ h \hat z \right) - \left( \bm v + w \hat z \right),
\end{equation}
where the first and second term give acoustic and optical contributions, respectively. From Eqs.\ \eqref{eq:Adef} and \eqref{eq:delta},
\begin{equation} \label{eq:A}
    \bm A = \frac{\sqrt{3}\hbar\beta}{2ea} \left[ \begin{pmatrix} u_{yy} - u_{xx} \\ u_{xy} + u_{yx} \end{pmatrix} + \frac{2 \sqrt{3}}{a} \hat z \times \left( \bm v + w \nabla h \right) \right],
\end{equation}
in lowest order with $\beta = -(a/\sqrt{3}t_0) \left. (\partial t/\partial d) \right|_0 \approx 3$ 
\cite{heeger_solitons_1988,castro_neto_electron-phonon_2007,guinea_gauge_2008} 
and $u_{ij} = \left[\partial_i u_j + \partial_j u_i + ( \partial_i h ) ( \partial_ j h ) \right]/2$ the strain tensor for acoustic displacements \cite{landau_theory_1986}. This expression only holds in the coordinate system shown in Fig.\ \ref{fig:fig1}(a). See SM for the general case \cite{Note1}. The first term in Eq.\ \eqref{eq:A} is the usual acoustic contribution ($\bm A_\text{ac}$) while the second term ($\bm A_\text{op}$) is new and gives a contribution from the relative motion between sublattices. This is one of the main results of this Letter. It is reminiscent of the renormalization of the electron-phonon coupling by optical phonons in pristine graphene \cite{woods_electron-phonon_2000,suzuura_phonons_2002}. In general, the optical component gives a pseudomagnetic field (PMF)
\begin{equation} \label{eq:Bop}
    B_\text{op} = \hat z \cdot \left( \nabla \times \bm A_\text{op} \right) = \frac{3 \hbar \beta}{ea^2} \nabla \cdot \left( \bm v + w \nabla h \right).
\end{equation}

To demonstrate our theory, we performed \textsc{lammps} molecular dynamics simulations for two graphene systems: moir\'e graphene and corrugated graphene. In all cases, we find that $\bm A_\text{op}$ acts to reduce the overall magnitude of the PMF.

\textcolor{NavyBlue}{\emph{Graphene moir\'es.}}---We first consider twisted bilayer graphene (TBG) \cite{lopesdossantos_graphene_2007,li_observation_2010,lopes_dos_santos_continuum_2012}, a twist moir\'e formed by stacking two layers of graphene with a relative angle, see Fig.\ \ref{fig:fig1}(c). In moir\'es, the atomic stacking between layers varies spatially. Certain stackings are favorable, and the systems relaxes to minimize the total elastic and adhesion energy \cite{carr_relaxation_2018}. This gives rise to atomic displacements and concomitant pseudogauge fields \cite{nam_lattice_2017}.

\begin{figure}
    \centering
    \includegraphics[width=\linewidth]{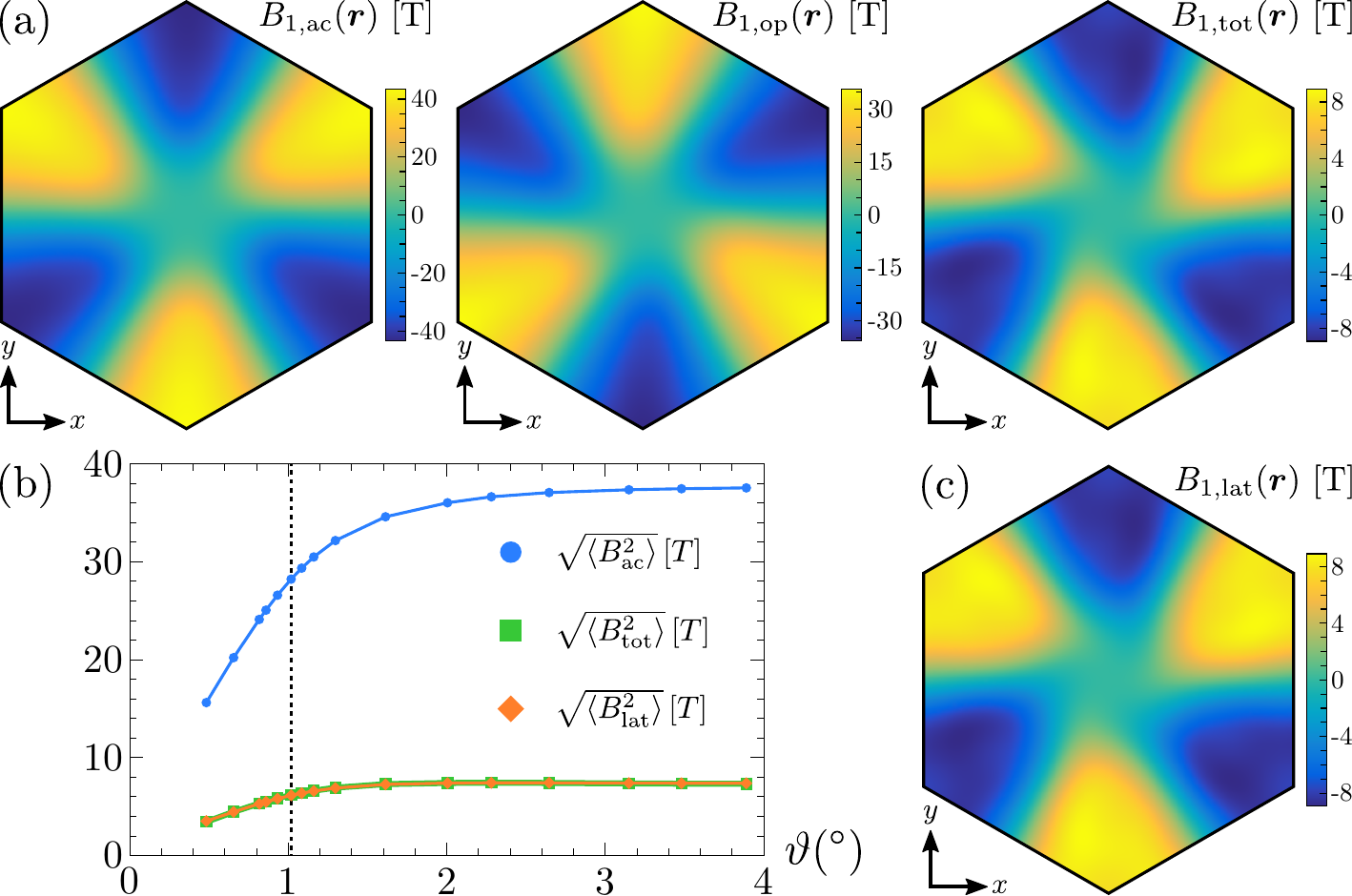}
    \caption{(a) PMF in the moir\'e cell from acoustic, optical, and total displacements due to lattice relaxation in TBG with twist angle $\vartheta \approx 1.018^\circ$. Shown for layer $1$ where layer $1,2$ is rotated counterclockwise by $\pm\vartheta/2$. Calculated from \textsc{lammps} simulations with $\beta = 3.37$ using $14$ reciprocal stars. Note that the $x$ axis corresponds here to the armchair direction. (b) RMS of the PMFs versus twist angle. (c) PMF calculated with Eq.\ \eqref{eq:Adef} for the hopping amplitude of Eq.\ \eqref{eq:hopping}.}
    \label{fig:fig2}
\end{figure}

As before, one defines displacement fields projected on the $xy$ plane, $\bm u_l$ and $\bm v_l$ with $l = 1,2$ the layer index, and similar for out-of-plane displacements. These displacements are calculated for a relaxed structure with \textsc{lammps} using the adaptive intermolecular reactive empirical bond order intralayer \cite{ stuart_reactive_2000} and dihedral-angle-corrected registry-dependent interlayer potential, benchmarked with state-of-the-art DFT \cite{leconte_relaxation_2022,Note1}. As expected for a twist moir\'e, the acoustic in-plane displacement field is almost entirely solenoidal, consistent with previous theory \cite{koshino_effective_2020,ezzi_analytical_2024} and experiment \cite{kazmierczak_strain_2021}. On the other hand, the optical field is mostly irrotational yielding a finite optical PMF \cite{Note1}. The acoustic, optical, and total PMF are shown in Fig.\ \ref{fig:fig2}(a). We only show the PMF for one layer since the $D_6$ symmetry of the moir\'e \cite{zou_band_2018} yields $\bm A_2(x,y) = \text{diag}(-1,1) \bm A_1(-x,y)$ from $\mathcal C_{2y}$ rotation symmetry. Moreover, the PMF is odd under $\bm r \mapsto -\bm r$ due to $\mathcal C_{2z}$ symmetry. Consequently, we have $B_2(x,y) = B_1(x,-y)$. We find that the acoustic and optical PMFs have similar shapes but opposite signs such that $B_\text{tot} = B_\text{ac} + B_\text{op}$ is about five times smaller than $B_\text{ac}$. We quantify this by plotting the root mean square (RMS) as a function of twist angle in Fig.\ \ref{fig:fig2}(b). The magnitudes of $B_\text{ac}$ and $B_\text{tot}$ are nearly constant above the magic angle \cite{ceferino_pseudomagnetic_2024,ezzi_analytical_2024} and the ratio $\sqrt{\left<B_\text{ac}^2\right>/\left<B_\text{tot}^2\right>} \approx 4.8$ for twists from $1^\circ$ to $4^\circ$. We also show $B_\text{lat}$ calculated from Eq.\ \eqref{eq:Adef} in Fig.\ \ref{fig:fig2}(c), using the atomic positions from \textsc{lammps} and intralayer hopping
\begin{equation} \label{eq:hopping}
    t(d) = t_0 \exp \left[ - \beta \left( \sqrt{3} d / a - 1 \right) \right].
\end{equation}
which depends only on the bond distance $d$ and $\beta = 3.37$ \cite{pereira_tight-binding_2009}. This is a good approximation because the nearest neighbors are approximately coplanar such that the overlap is dominated by $V_{pp\pi} = -t_0$.

Our results show that optical displacements, i.e., the relative motion of different sublattices, effectively screen the PMF in twisted graphene, yielding much lower values as previously predicted \cite{nam_lattice_2017}. Physically, we find that relative displacements act to restore the microscopic $\mathcal C_{3z}$ symmetry by reducing changes in the bond length.

\textcolor{NavyBlue}{\emph{Corrugated graphene.}}---As a second example, we consider monolayer graphene subjected to a long-wavelength corrugation, see Fig.\ \ref{fig:fig1}(c). This setup may be realized by engineering a suitable substrate \cite{jiang_visualizing_2017,kang_pseudo-magnetic_2021,phong_boundary_2022,zhang_electronic_2018,zhang_magnetotransport_2019} or through a buckling transition as was observed for graphene on NbSe$_2$ \cite{mao_evidence_2020,milovanovic_band_2020}. In particular, we consider a periodic corrugation with $C_{3v}$ symmetry \cite{de_beule_network_2023,de_beule_roses_2023,gao_untwisting_2023,mahmud_topological_2023} commensurate with graphene and given by $h_\text{sub}(\bm r) = h_0 \sum_{n=1}^3 \cos \left( \G_n \cdot \bm r + \theta \right)$ where $\bm g_1 = 4\pi / ( \sqrt{3} L) \left( 0, 1 \right)$ and $\bm g_{2,3} = 4\pi / ( \sqrt{3} L ) \left( \mp \sqrt{3}/2, -1/2 \right)$. These are three of the shortest nonzero reciprocal vectors of the corrugation related by $\mathcal C_{3z}$ rotations. Here $h_0$ and $\theta$ control the amplitude and shape of the corrugation and $L = Na$ is the superlattice constant with integer $N\gg1$.

The corrugated substrate is modeled in \textsc{lammps} with a dense honeycomb lattice to ensure commensuration and smoothness. For interactions in the graphene, we use the standard reactive empirical bond order potential \cite{stuart_reactive_2000}. The interaction with a generic substrate in the absence of moir\'e effects is modeled with a $12$-$6$ Lennard-Jones potential \cite{Note1}. These capture the binding energy with the substrate, which can be used as a tuning parameter. The resulting in-plane displacements are analyzed using a Helmholtz decomposition, $\bm u = \sum_{\bm g} ( u_{\bm g}^\parallel \bm g + u_{\bm g}^\perp \hat z \times \bm g ) e^{i\bm g \cdot \bm r} / (ig^2)$ and similar for $\bm v$. Here the longitudinal and transverse coefficients $u_{\bm g}^\parallel$ and $u_{\bm g}^\perp$ are c-numbers corresponding to the curl and divergence, respectively. These coefficients are constrained by symmetry \cite{ezzi_analytical_2024,Note1} and together with $h_{\bm g}$ and $w_{\bm g}$ give four complex coefficients for each reciprocal star. For example, for $\theta = 0$ (modulo $\pi/3$) the corrugation has $C_{6v}$ symmetry. In this case, $\mathcal C_{2z}$ implies $\bm u(\bm r) = -\bm u(-\bm r)$ and $h(\bm r) = h(-\bm r)$ but $\bm v(\bm r) = \bm v(-\bm r)$ and $w(\bm r) = -w(-\bm r)$ since a $\mathcal C_{2z}$ rotation exchanges the sublattices. 

As a first approximation, we use continuum elasticity for the acoustic in-plane displacement field. In the limit where graphene is pinned to the substrate: $h=h_\text{sub}$ \cite{guinea_gauge_2008} and the only nonzero in-plane coefficients are $u_1^\parallel = (1 - 3\nu) \pi^2 h_0^2 e^{-2i\theta} / (3L^2)$, $u_2^\parallel = (3-\nu) \pi^2 h_0^2/ (3L^2)$, and $u_3^\parallel = 2\pi^2 h_0^2  e^{2i\theta}/ (3L^2)$ where the subscript indexes the star and $\nu \approx 0.165$ is the Poisson ratio \cite{lee_measurement_2008}. From \textsc{lammps}, we also find that volumetric components are dominant for both $\bm u$ and $\bm v$. While some rotational components are symmetry-allowed, they are at least one order of magnitude smaller \cite{Note1}. Thus, unlike twist moir\'es, $\bm u$ is mostly irrotational for corrugations. 

\begin{figure}
    \centering
    \includegraphics[width=\linewidth]{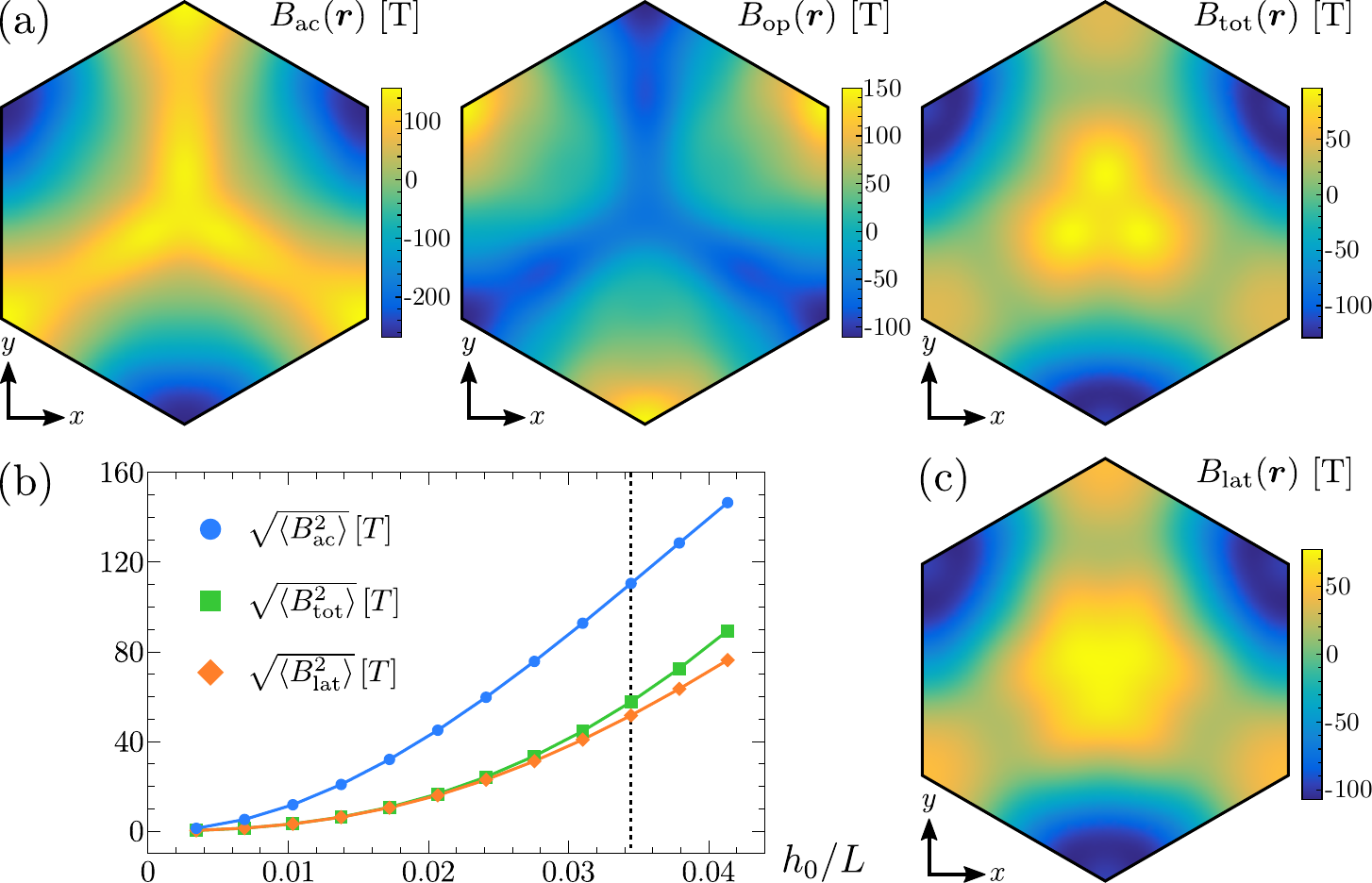}
    \caption{PMF in the supercell from acoustic ($B_\text{ac}$), optical ($B_\text{op}$), and total ($B_\text{tot}$) displacements for a $C_{3v}$ periodic height corrugation with $L=59a$, $h_0 = 5 \, \text{\r A}$, and $\theta = 15^\circ$. Calculated from \textsc{lammps} simulations with $\beta = 3.37$ using $14$ reciprocal stars. (b) RMS of the PMFs for $\theta = 15^\circ$ versus $h_0/L$. The dashed line indicates the value for (a) and (c). (c) PMF calculated from tight binding through Eqs.\ \eqref{eq:Adef} and \eqref{eq:hopping}.}
    \label{fig:fig3}
\end{figure}

As was the case for TBG, optical displacements are significant even in the regime where continuum elasticity accurately describes the acoustic displacements ($h_0/L < 0.02$). While generally $\bm v$ is at least one order of magnitude smaller than $\bm u$, the former contributes to the pseudogauge field at zeroth order and the latter via the strain tensor, see Eq.\ \eqref{eq:A}. Hence the acoustic part is suppressed by a factor $a/L$ such that both fields contribute at the same order. In Fig.\ \ref{fig:fig3}(a) we show the PMFs for a corrugation with $h_0 = 5 \, \text{\r A}$ and $\theta = 15^\circ$. We see that the total PMF is more concentrated and its magnitude is halved compared to $B_\text{ac}$. This is further illustrated in Fig.\ \ref{fig:fig3}(b) where the RMS of the PMFs is plotted as a function of $h_0/L$. For $h_0/L < 0.02$, the PMF from Eq.\ \eqref{eq:A} matches $B_\text{lat}$ from Eq.\ \eqref{eq:Adef} directly using Eq.\ \eqref{eq:hopping} with \textsc{lammps} results, only if we include optical displacements. For large amplitudes $h_0/L > 0.02$, there are higher-order corrections which further reduce the magnitude of the PMF.

\begin{figure}
    \centering
    \includegraphics[width=\linewidth]{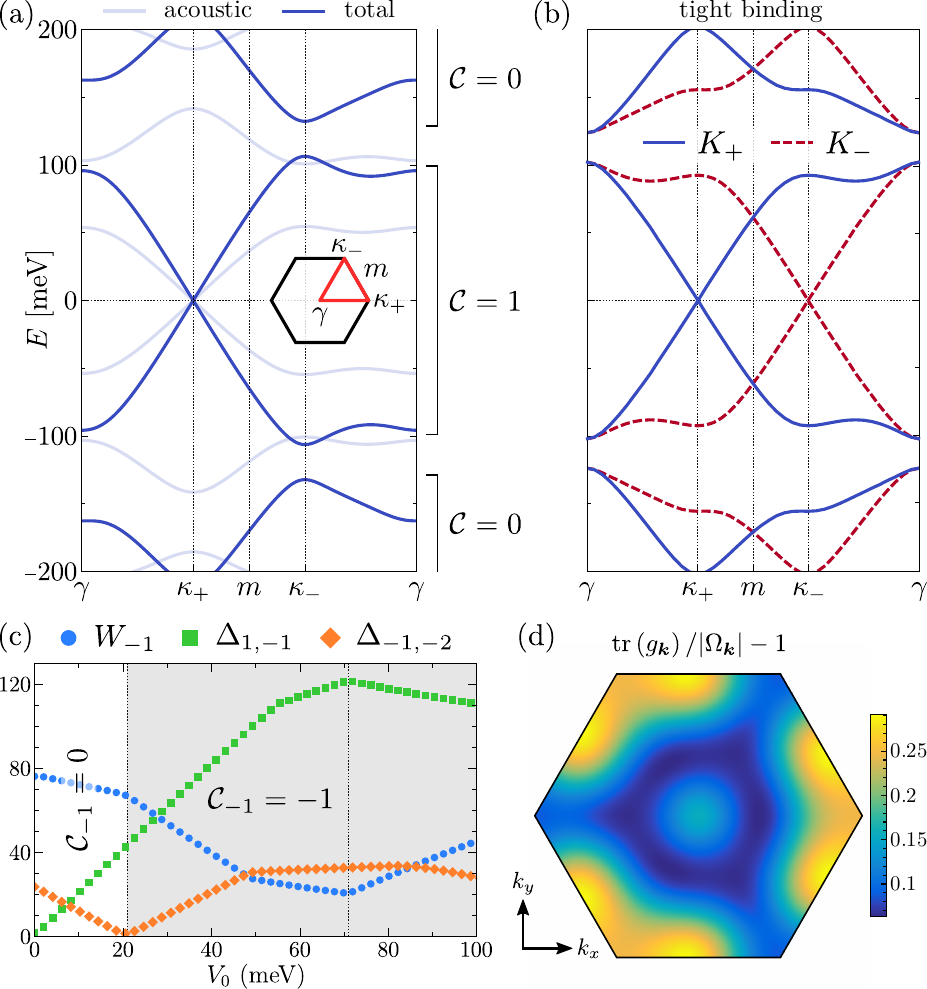}
    \caption{Electronic bands near charge neutrality for graphene subject to a periodic $C_{3v}$ corrugation with period $L=59a\approx14.5 \, \text{nm}$, $h_0 = 5 \, \text{\r A}$, and $\theta = 15^\circ$.  (a) Continuum model for valley $K_+$ with only acoustic (light) and total (dark) displacements. Plotted along high-symmetry lines (inset). The other valley is related by time reversal and the valley Chern number $\mathcal C$ is indicated. (b) Tight binding using atomic positions from \textsc{lammps} where the valley polarization is calculated with a modified Haldane hopping \cite{Note1}. (c) Bandwidth $W_n$ and gaps $\Delta_{n,n-1}$ for the first valence band ($n=-1$) for $h_0 = 6 \, \text{\r A}$ and $\theta = 15^\circ$ as a function of an electric field along the $z$ direction, showing the valley Chern number. Kinks in the topological region are due to direct-to-indirect gap transitions. (d) Violation of trace condition for the minimum bandwidth in (c).}
    \label{fig:fig4}
\end{figure}

The suppression of the pseudogauge field due to optical displacements strongly modifies the electronic minibands near charge neutrality. In Fig.\ \ref{fig:fig4}, we show the bands calculated with the continuum model together with those from the tight-binding model in Eq.\ \eqref{eq:tb}, implemented in the \textsc{pybinding} software \cite{moldovan_pybinding_2020}, using atomic positions from \textsc{lammps} with the hopping amplitude from Eq.\ \eqref{eq:hopping}. As expected, optical contributions reduce the minigaps between these bands and increase the bandwidth, while the topology of the bands remains unchanged. The latter is given by the valley Chern number $\mathcal C = \left( \mathcal C_+ - \mathcal C_- \right) / 2$ since the magnetic point group in a single valley breaks time-reversal symmetry. For a $C_{3v}$ corrugation, it is given by $C_{3v}(C_3) = \left< \mathcal C_{3z}, \mathcal M_x \mathcal T \right>$ where $\mathcal M_x (x \mapsto -x)$ is an in-plane mirror and $\mathcal T$ is spinless time reversal \cite{dresselhaus_group_2007}. Importantly, the band manifold near charge neutrality is well reproduced by the continuum model only if we include optical contributions. However, the continuum model fails to reproduce the remote bands. This discrepancy is likely due to higher-order corrections in the continuum theory \cite{ramezani_masir_pseudo_2013} such as a position-dependent Fermi velocity \cite{pellegrino_transport_2011,de_juan_space_2012,de_juan_gauge_2013,amorim_novel_2016}. Moreover, using Eq.\ \eqref{eq:Adef} directly by taking its Fourier transform changes only slightly the continuum bands.

The reduced band flattening and gaps from the elastically screened gauge field, has important consequences for the feasibility of symmetry-broken phases \cite{manesco_correlations_2020,manesco_correlation-induced_2021} and fractional Chern insulators \cite{gao_untwisting_2023} in periodically corrugated graphene. Nevertheless, in the presence of an electric field normal to the $xy$ plane, which couples to the height modulation \cite{gao_untwisting_2023}, one can still obtain isolated and flattened minibands. This results from the sublattice polarization near charge neutrality induced by the PMF in real space \cite{georgi_tuning_2017,mao_evidence_2020} such that a scalar potential $V(\bm r) = V_0 h(\bm r) / h_0$ effectively acts as a staggered sublattice potential on the superlattice scale. This is shown in Fig.\ \ref{fig:fig4}(c) where we plot the bandwidth $W$ and gaps $\Delta$ of the highest valence band versus $V_0$ for $h_0 = 6 \, \text{\r A}$ and $\theta = 15^\circ$. As the gap opens at the Dirac point, the 2nd valence band closes and reopens to a topological phase with a minimum bandwidth of $20\, \text{meV}$ for a field strength $40 \, \text{mV}/\text{nm}$. Moreover, the violation of the trace condition $\text{tr}(g_{\bm k}) = |\Omega_{\bm k}|$ with $g_{\bm k}$ the quantum metric and $\Omega_{\bm k}$ the Berry curvature, is on the order of 10\%, see Fig.\ \ref{fig:fig4}(d). This quantifies how well a $|\mathcal C| = 1$ Bloch band mimics a Landau level (for which the equality holds) \cite{roy_band_2014}. Hence, the reduced sublattice polarization from the screened PMF makes this system less favorable for hosting a fractional Chern insulator \cite{gao_untwisting_2023}.

\textcolor{NavyBlue}{\emph{Conclusions.}}---We developed a theory of pseudogauge fields in graphene that takes into account contributions from both acoustic and optical displacement fields. The latter corresponds to the relative motion of different sublattices. Using molecular dynamics simulations, we have shown that optical displacements significantly modify the resulting pseudomagnetic field. Specifically, we applied our theory to moir\'e graphene and corrugated graphene. In all cases studied, optical contributions screen the acoustic contribution resulting in an overall reduction of the pseudomagnetic field. A simple explanation is given by that fact that the internal relaxation tends to restore bond lengths to their pristine value.

Our theory elucidates the origin of discrepancies between continuum and tight-binding calculations that use microscopic theories to model lattice relaxation. It also introduces a novel way to engineer pseudomagnetic fields through the optical displacement, though this may be difficult to achieve in practice and requires microscopic theories that go beyond continuum elasticity. Furthermore, our theory will help to understand experiments that probe pseudogauge fields. For example, in twisted bilayer graphene, elastic screening of pseudomagnetic fields may explain why initial predictions based on continuum elasticity \cite{nam_lattice_2017} were not observed.

We conclude that continuum elasticity, which only yields the acoustic displacement field, cannot fully describe pseudogauge fields in graphene and most likely other low-dimensional materials with nonprimitive lattices, e.g., transition metal dichalcogenides. For displacements that mostly lie in the nominal graphene plane, a reduction factor \cite{woods_electron-phonon_2000,suzuura_phonons_2002} may be used for qualitative results. However, the reduction factor is not universal and may vary depending on the amount of strain and microscopic details. Moreover, when out-of-plane displacements are significant, a reduction factor is inadequate even for qualitative results because both the magnitude and shape of the pseudomagnetic field are modified.

\let\oldaddcontentsline\addcontentsline 
\renewcommand{\addcontentsline}[3]{} 
\begin{acknowledgments}
CDB and EJM are supported by the U.S.\ Department of Energy under Grant No.\ DE-FG02-84ER45118. RS, WNL and LC acknowledge support from Research Foundation-Flanders (FWO) research project No.\ G0A5921N and the EOS project ShapeME. The computational resources used in this work were provided by the HPC core facility CalcUA of the University of Antwerp, and the Flemish Supercomputer Center (VSC), funded by FWO and the Flemish Government. We thank Shaffique Adam and Mohammed M.\ Al Ezzi for interesting and helpful discussions. We further acknowledge Gayani N.\ Pallewela for sharing \textsc{lammps} data of twisted bilayer graphene for different molecular dynamics potentials, and Bert Jorissen for the implementation of the valley operator in the \textsc{pybinding} software.
\end{acknowledgments}

\bibliography{references}
\let\addcontentsline\oldaddcontentsline 


\clearpage
\onecolumngrid
\begin{center}
\textbf{\Large Supplemental Material for ``Elastic Screening of Pseudogauge Fields in Graphene''}
\end{center}
 
\setcounter{equation}{0}
\setcounter{figure}{0}
\setcounter{table}{0}
\setcounter{page}{1}
\setcounter{secnumdepth}{2}
\makeatletter
\renewcommand{\thepage}{S\arabic{page}}
\renewcommand{\thesection}{S\arabic{section}}
\renewcommand{\theequation}{S\arabic{equation}}
\renewcommand{\thefigure}{S\arabic{figure}}
\renewcommand{\thetable}{S\arabic{table}}

\tableofcontents
\vspace{1cm}

\twocolumngrid

\section{Electronic theory} \label{app:electronic}

\subsection{Continuum limit}

The tight-binding Hamiltonian of graphene in the nearest-neighbor approximation in the presence of strain can be written as
\begin{equation}
    H = -\sum_{\bm r} \sum_{n=1}^3 t_n(\bm r) c_A^\dag(\bm r) c_B\left( \bm r + \bm \delta_n^0 \right) + \text{h.c.}, 
\end{equation}
where the first sum runs over cells $\bm r$ and the second one over nearest neighbors. Note we use the original pristine positions to label the sites. Here $t_n(\bm r) > 0$ is the hopping amplitude between nearest neighbors that is modulated by strain, and $c_\sigma^\dag(\bm r)$ [$c_\sigma(\bm r)$] are creation (annihilation) operators for sublattice $\sigma = A,B$. The position of $A$ atoms is given by $\bm r + \bm u_A(\bm r) + h_A(\bm r) \hat z$ where $\bm u_A$ is the in-plane displacement, i.e, projected on the original graphene $xy$ plane, and $h_A$ is the out-of-plane displacement in the $z$ direction, and similar for $B$ atoms. 

\begin{figure}
    \centering
    \includegraphics[width=.6\linewidth]{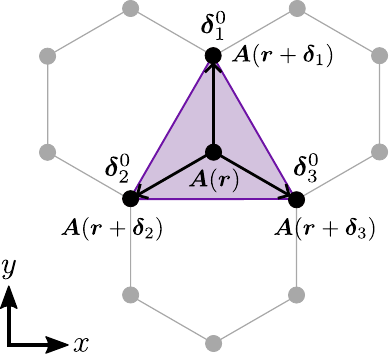}
    \caption{Pristine graphene lattice with the zigzag direction along the $x$ axis. The three nearest-neighbor bonds are shown together with an illustration of the calculation of the PMF in the central atomic position using discrete contour integration.}
    \label{fig:graphene_app}
\end{figure}

To study the low-energy physics near the $\bm K_\pm$ point, we take the continuum limit. This amounts to the replacement
\begin{equation} \label{eq:spinor}
    c_\sigma(\bm r) \rightarrow \sqrt{A_c} \sum_\tau \psi_{\tau \sigma}(\bm r) e^{i \bm K_\tau \cdot \bm r},
\end{equation}
where $A_c = A/N$ is the unit cell area. Here, $\psi_{\tau\sigma}^\dag(\bm r)$ [$\psi_{\tau\sigma}(\bm r)$] are field operators that create (annihilate) a fermion of sublattice $\sigma$ at position $\bm r$ composed of small momentum components $|\bm q|a \ll 1$ near valley $\bm K_\tau$ with $\tau = \pm 1$ the valley index, and which obey the usual fermionic relations. Note that we evaluate the field operator at the unperturbed position, which amounts to neglecting ``frame effects'' \cite{de_juan_gauge_2013}. Such effects can be included by letting $\bm r \rightarrow \bm r + \bm u_{\sigma}(\bm r)$ in \eqref{eq:spinor}. However, since the change in the phase is slowly varying on the lattice scale it can always be removed by absorbing it into the spinor: $\psi_\sigma(\bm r) \rightarrow \psi_\sigma(\bm r) e^{-i \bm K_\tau \cdot \bm u_\sigma(\bm r)}$. Thus observable frame effects only appear at second order. Here we are only interested in the lowest order such that we can safely neglect them. The effective Hamiltonian becomes
\begin{widetext}
\begin{align}
    H_\text{eff} & = -\sum_\tau \sum_{n=1}^3 \int d^2 \bm r \, t_n(\bm r) e^{i \bm K_\tau \cdot \bm \delta_n^0} \psi_{\tau A}^\dag(\bm r) \psi_{\tau B} \left( \bm r + \bm \delta_n^0 \right)  + \text{h.c.} \\ 
    & \approx -\sum_\tau \int d^2 \bm r \, \psi_{\tau A}^\dag(\bm r) \sum_{n=1}^3 e^{i \bm K_\tau \cdot \bm \delta_n^0} \left[ t_0 \bm \delta_n^0 \cdot \nabla_{\bm r} + \delta t_n(\bm r) \right] \psi_{\tau B}(\bm r) + \text{h.c.},
\end{align}
\end{widetext}
where we let $\sum_{\bm r} \rightarrow A_c^{-1} \int d^2 \bm r$. In the second line, we expanded everything up to lowest order in gradients and displacements with $t_n(\bm r) = t_0 + \delta t_n(\bm r)$, where we take $t_0 = 2.8$~eV \cite{castro_neto_electronic_2009}. We further defined the nearest-neighbor bond vectors of pristine graphene $\bm \delta_n^0$. For example, for the orientation shown in Fig.\ \ref{fig:graphene_app}, we have $\bm \delta_1^0 = \left( 0, a_0 \right)$, $\bm \delta_2^0 = a_0 \left( -\sqrt{3}/2, -1/2 \right)$, and $\bm \delta_3^0 = a_0 \left( \sqrt{3}/2, -1/2 \right)$ where $a_0 = 1.42 \; \text{\r A}$ is the nearest-neighbor distance.

Moreover, we have assumed that intervalley coupling is negligible. This is justified in the limit $L \gg a$ where the strain field varies slowly with respect to the graphene lattice. In the following, we consider a general orientation of the graphene where $\varphi$ is the angle between the zigzag direction and the $x$ axis. For example, we have $\varphi = 0$ for zigzag orientation, shown in Fig.\ \ref{fig:graphene_app}, and armchair orientation would correspond to $\varphi = -\pi/2$. We now take $\bm K_\pm = \pm R(\varphi) (4\pi/3a, 0)$ with $R(\varphi)$ the standard $2 \times 2$ rotation matrix for a counterclockwise rotation by an angle $\varphi$. One finds \cite{katsnelson_graphene_2007,vozmediano_gauge_2010}
\begin{equation}
    -t_0 \sum_{n=1}^3 e^{i \bm K_\tau \cdot \bm \delta_n^0} \bm \delta_n^0 \cdot \nabla_{\bm r} = -i \hbar v_F  e^{i\tau \varphi} \left( \tau \partial_x - i \partial_y \right),
\end{equation}
with $\hbar v_F = \sqrt{3} t_0 a/2$ and we define the vector potential $\bm A = (A_x, A_y)$ as
\begin{equation}
    -\sum_{n=1}^3 \delta t_n(\bm r) e^{i \bm K_\tau \cdot \bm \delta_n^0} \equiv e v_F e^{i\tau \varphi} \left[ A_x(\bm r) - i \tau A_y(\bm r) \right],
\end{equation}
with $-e$ the electron charge. Explicitly,
\begin{equation} \label{eq:pvp}
    \bm A(\bm r) = \frac{R(\varphi)}{2ev_F} \begin{bmatrix}  \delta t_2 + \delta t_3 - 2 \delta t_1 \\
    \sqrt{3} \left( \delta t_3 - \delta t_2 \right)
    \end{bmatrix}.
\end{equation}
The effective low-energy Hamiltonian thus becomes
\begin{widetext}
\begin{equation} \label{eq:Heffapp}
    H_\text{eff} = \hbar v_F \sum_\tau \int d^2 \bm r \, \psi_\tau^\dag(\bm r) \left\{ \left[ -i \nabla_{\bm r} + \frac{e\tau}{\hbar} \bm A(\bm r) \right] \cdot (\tau \sigma_x, \sigma_y) \right\} \psi_\tau(\bm r),
\end{equation}
with $\psi_\tau(\bm r) = \left[ e^{-i\tau \varphi/2} \psi_{\tau A}(\bm r), e^{i\tau \varphi/2} \psi_{\tau B}(\bm r) \right]^t$. 

To compute the change in hopping amplitude $\delta t_n$ due to strain, we first consider the change in the nearest-neighbor bond vectors. In lowest order of displacements and their gradients, we have
\begin{align}
    \bm \delta_n(\bm r)- \bm \delta_n^0 & = \bm u_B(\bm r + \bm \delta_n^0) - \bm u_A(\bm r) + [ h_B(\bm r + \bm \delta_n^0) - h_A(\bm r) ] \hat z \\
    & \approx \left( \bm \delta_n^0 \cdot \nabla \right) \left[ \bm u(\bm r) + h(\bm r) \hat z \right] - \left[ \bm v(\bm r) + w(\bm r) \hat z \right],
\end{align}
\end{widetext}
where
\begin{alignat}{2}
    \bm u & = ( \bm u_A + \bm u_B ) / 2, \qquad && h = ( h_A + h_B ) / 2, \\
    \bm v & = \bm u_A - \bm u_B, \qquad && w = h_A - h_B,
\end{alignat}
are the center-of-mass (acoustic) and relative (optical) displacements, respectively. For example, in a classical microscopic theory of the in-plane phonon modes of a \emph{pristine} graphene sheet, one can show that in the long-wavelength limit \cite{woods_electron-phonon_2000,suzuura_phonons_2002}
\begin{equation}
    \bm v(\bm r) = \frac{\left( \kappa - 1 \right) a}{2\sqrt{3}} \, R(3\varphi) \begin{pmatrix} \partial_x u_y + \partial_y u_x \\ \partial_x u_x - \partial_y u_y \end{pmatrix},
\end{equation}
with $\kappa \sim 1/3$ the so-called reduction factor, whose precise value depends on microscopic details. We do not attempt to find a relation between the optical and acoustic displacements for the cases we consider, e.g., moir\'e graphene and corrugated graphene. Instead, our microscopic theory is given by a molecular dynamics simulation which yields the fields $\bm u_{A,B}$ and $h_{A,B}$. 

Next we calculate the change in the hopping amplitude. To this end, we assume that $t(\bm d) = t(d)$ only depends on the bond distance $d$ (two-center approximation) and expand them in lowest order of the displacements:
\begin{widetext}
\begin{align}
    \delta t_n & = t( \bm \delta_n ) - t_0 \\
    & \approx \left. \frac{\partial t}{\partial d_i} \right|_0 \left( \bm \delta_n - \bm \delta_n^0 \right)_i + \frac{1}{2} \left. \frac{\partial^2 t}{\partial d_z^2} \right|_0 \left( \bm \delta_n - \bm \delta_n^0 \right)_z \left( \bm \delta_n - \bm \delta_n^0 \right)_z \\
    & = -\frac{3\beta t_0}{a^2} \left[ \bm \delta_n^0 \cdot \bm \left( \bm \delta_n - \bm \delta_n^0 \right) + \frac{1}{2} \left( \bm \delta_n - \bm \delta_n^0 \right)_z \left( \bm \delta_n - \bm \delta_n^0 \right)_z \right], \\
    & \equiv -\frac{3\beta t_0}{a^2} \left[ \delta_{ni}^0 \delta_{nj}^0 u_{ij} - \delta_{ni}^0\left( v_i + w \partial_i h \right) + \frac{w^2}{2} \right],
\end{align}
where $\beta = -\frac{a}{\sqrt{3}t_0} \left. \frac{\partial t}{\partial d} \right|_{\text{nn},0} \approx 3$ \cite{heeger_solitons_1988,castro_neto_electron-phonon_2007,guinea_gauge_2008,castro_neto_electronic_2009,low_gaps_2011}, $u_{ij}$ is the strain tensor, and we used that $d_z = 0$ in the absence of strain. We further used ($i = x, y$)
\begin{align}
    \frac{\partial t}{\partial d_i} & = \frac{d_i}{d} \frac{\partial t}{\partial d}, \\
    \frac{\partial t}{\partial d_z} & = \frac{d_z}{d} \frac{\partial t}{\partial d}, \\
    \frac{\partial^2 t}{\partial d_z^2} & = \frac{1}{d} \frac{\partial t}{\partial d} - \frac{d_z^2}{d^3} \frac{\partial t}{\partial d} + \frac{d_z^2}{d^2} \frac{\partial t^2}{\partial d^2}, \\
    \frac{\partial^2 t}{\partial d_z \partial d_i} & = \frac{d_i d_z}{d^2} \frac{\partial^2 t}{\partial d^2} - \frac{d_i d_z}{d^3} \frac{\partial t}{\partial d}.
\end{align}

Plugging the result for $\delta t_n$ into Eq.\ \eqref{eq:pvp}, we obtain
\begin{equation}
    \bm A = \bm A_\text{ac} + \bm A_\text{op} = \frac{\sqrt{3}\hbar\beta}{2ea} \left[ R(3\varphi) \begin{pmatrix} u_{yy} - u_{xx} \\ u_{xy} + u_{yx} \end{pmatrix} + \frac{2 \sqrt{3}}{a} \hat z \times \left( \bm v + w \nabla h \right) \right],
\end{equation}
which is preserved under a global $\mathcal C_{3z}$ rotation, as both the graphene lattice and the valley are left unchanged. 
\end{widetext}

For completeness, we also consider the second nearest-neighbor hopping $t_m'(\bm r)$ between atoms of the same sublattice. This yields an additional term $V(\bm r) \psi_\tau^\dag(\bm r) \psi_\tau(\bm r)$ in the effective Hamiltonian of Eq.\ \eqref{eq:Heff} where $V(\bm r)$ is the deformation potential. In lowest order of the displacements and up to a constant energy shift, we find 
\begin{align}
    V(\bm r) & = -\sum_{m=1}^6 \delta t_m'(\bm r) e^{i \bm K_\tau \cdot \bm \eta_m^0} \\
    & = \delta t_1' + \delta t_2' + \delta t_3' \\
    & = \frac{3a}{2} \left. \frac{\partial t}{\partial d} \right|_{\text{nnn},0} \left( u_{xx} + u_{yy} \right), \label{eq:PSFnnn}
\end{align}
where $\bm \eta_1^0 = \bm a_1$, $\bm \eta_2^0 = \bm a_2$, $\bm \eta_3^0 = -\bm a_1 - \bm a_2$, $\bm \eta_4^0 = -\bm a_1$, $\bm \eta_5^0 = -\bm a_2$, and $\bm \eta_6^0 = \bm a_1 + \bm a_2$ are the six second nearest-neighbor bond vectors of pristine graphene with $\bm a_{1/2} = \bm \delta_1^0 - \bm \delta_{2/3}^0$ the primitive lattice vectors. Here we also used the two-center approximation such that $\delta t'$ is independent on reversal of the bond vector. We can estimate the prefactor in front of $\text{tr}(u) = u_{xx} + u_{yy}$ in Eq.\ \eqref{eq:PSFnnn} with a simple model for the hopping amplitude,
\begin{equation}
    t(d) = t_0 \exp \left[ - \beta \left( d / a_0 - 1 \right) \right],
\end{equation}
which yields a prefactor $-3 \sqrt{3} \beta t(a) / 2$. The deformation potential has full rotational symmetry and is finite even in the presence of microscopic $\mathcal C_{3z}$ such as for biaxial strain \cite{van_der_donck_piezoelectricity_2016}. Moreover, it does not depend on the relative displacements in lowest order because it couples equal sublattices.

\subsection{Tight binding} \label{sec:TB}

In this section, we describe the tight-binding method used to calculate the electronic properties. The advantage of this approach is that it allows us to construct a model directly from the atomic configurations obtained from the molecular dynamics simulations, fully taking into account the microscopic details of the lattice relaxation. The tight-binding Hamiltonian is given by
\begin{equation}
    H = -\sum_{\left< i, j \right>} t_{ij} c_i^\dag c_j + \sum_i \xi_i c_i^\dag c_i,
    \label{eq:tb_hamiltonian}
\end{equation}
where the atomic sites are labeled by $i$ and $j$ and the first sum runs over nearest neighbors. The nearest-neighbor hopping amplitude $t_{ij}$ is approximated as
\begin{equation}
    t_{ij} = t_0 \exp{\left[- \beta \left( \frac{|\bm r_i - \bm r_j|}{a_0} - 1 \right) \right]},
\end{equation}
where we use the values $t_0 = 2.8$~eV, $a_0 = a / \sqrt{3} = 0.142$~nm, and $\beta = 3.37$ \cite{castro_neto_electronic_2009,pereira_tight-binding_2009}. This is expected to be a good approximation since the nearest-neighbors lie approximately in the same plane, which generally differs from the $xy$ plane in the presence of corrugation. Hence, the relevant overlap integral is still given by $V_{pp\pi} = t_0$.

The second term in Eq.\ \eqref{eq:tb_hamiltonian} is an on-site electrostatic potential, which can originate from an external electric field or the pseudo scalar field \cite{grassano_work_2020, choi_effects_2010}. 

\subsection{Pseudomagnetic field on a discrete grid} \label{sec:pmf_discrete}

We can further calculate the pseudomagnetic field (PMF) directly from the atomic positions, allowing us to determine the PMF that effectively enters the tight-binding calculations. For the zigzag orientation shown in Fig.\ \ref{fig:graphene_app}, the vector potential is given by
\begin{equation} \label{eq:Adef_supp}
    A_x(\bm r) - i A_y(\bm r) = -\frac{1}{e v_F}\sum_{n=1}^3 \delta t_n(\bm r) e^{i\bm K \cdot \bm \delta_n(\bm r)},
\end{equation}
where $\delta t_{n} = t_{ij} - t_{0}$ is the change in hopping energy, $\bm K_+ = 4 \pi / (3a) \hat{x}$ the zone corner of the graphene BZ and 
\begin{equation}
    \begin{aligned}
        \bm \delta_n(\bm r) = \bm \delta_{n}^0 & + \bm u_B(\bm r + \bm \delta_n^0) + h_B(\bm r + \bm \delta_n^0) \hat z \\
        & - \bm u_A(\bm r) - h_A(\bm r) \hat z,
    \end{aligned}
\end{equation}
the modified bond vector. Please note that the definition in Eq.\ \eqref{eq:Adef_supp} differs from our previous definition since we also take into account changes in the phase factor. These only modify the PMF at next-to-leading order as the lowest order contribution from the phase factor corresponds to a gauge transformation. We prefer this definition for this section since all contributions to \eqref{eq:Adef_supp} are automatically taken into account in tight-binding calculations, whereas in the continuum theory we only consider leading-order terms from an expansion in displacements and momentum \cite{ramezani_masir_pseudo_2013}.

The PMF is given by $\bm B  = \nabla \times \bm A$ and calculated on a discrete (strained) atomic grid using Stokes' theorem. In two dimensions, this relates the surface integral of the curl of a vector field $\bm A$ to the contour integral around the boundary of the same field:
\begin{equation}
    \iint_\Sigma (\nabla \times \bm A) \cdot d\bm \Sigma = \oint_{\partial \Sigma} \bm A \cdot d \bm r.
    \label{eq:stokes}
\end{equation}
If we consider $\bm B$ to be slowly varying within the graphene unit cell, the surface integral in \eqref{eq:stokes} becomes trivial:
\begin{align}
    \nabla \times \bm A(\bm r) & \approx \frac{1}{S} \oint_{\partial \Sigma} \bm A \cdot d \bm r \label{eq:Alat} \\ 
    & = \frac{1}{S} \left( \int_{\bm r + \bm \delta_1}^{\bm r + \bm \delta_2} + \int_{\bm r + \bm \delta_2}^{\bm r + \bm \delta_3} + \int_{\bm r + \bm \delta_3}^{\bm r + \bm \delta_1} \right) \bm A \cdot d \bm r \nonumber \\
    & \approx \frac{1}{S} \left[ \bm A(\bm r + \bm \delta_1) + \bm A(\bm r + \bm \delta_2) \right] \cdot \frac{\bm \delta_2 - \bm \delta_1}{2} \nonumber \\ 
    & + \frac{1}{S} \left[ \bm A(\bm r + \bm \delta_2) + \bm A(\bm r + \bm \delta_3) \right] \cdot \frac{\bm \delta_3 - \bm \delta_2}{2} \nonumber \\ 
    & + \frac{1}{S} \left[ \bm A(\bm r + \bm \delta_3) + \bm A(\bm r + \bm \delta_1) \right] \cdot \frac{\bm \delta_1 - \bm \delta_3}{2} \nonumber \\
    & = \bm A(\bm r + \bm \delta_1) \cdot \frac{\bm \delta_2 - \bm \delta_3}{2S} + \bm A(\bm r + \bm \delta_2) \cdot \frac{\bm \delta_3 - \bm \delta_1}{2S} \nonumber \\ 
    & + \bm A(\bm r + \bm \delta_3) \cdot \frac{\bm \delta_1 - \bm \delta_2}{2S}, \nonumber
\end{align}
where we approximated the line integral along the triangle contour shown in Fig.\ \ref{fig:graphene_app}. The enclosed area $S$ is calculated by taking the cross product of any two sides,
\begin{equation}
    S = \frac{|(\bm \delta_1 - \bm \delta_2) \times (\bm \delta_3 - \bm \delta_2)|}{2}.
\end{equation}
Finally, we note that this formula is only gauge invariant up to the same order of approximation. In particular, if we send $\bm A \mapsto \bm A + \nabla \chi$ and approximate $\nabla \chi(\bm r+\bm \delta_n) \simeq \nabla \chi(\bm r) + \left( \bm \delta_n \cdot \nabla \right) \nabla \chi(\bm r)$ one can show that the final expression in Eq.\ \eqref{eq:Alat} remains unchanged.

\section{Moir\'e graphene} \label{app:moire}

\begin{figure}
    \centering
    \includegraphics[width=.6\linewidth]{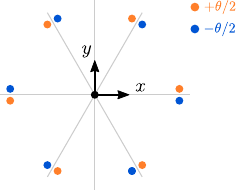}
    \caption{Twisted bilayer graphene with $D_6$ symmetry showing only carbon atoms closest to the twist center. Symmetry axes are shown in gray.} 
    \label{fig:tbgD6}
\end{figure}

In this section, we discuss the displacement fields from lattice relaxation in twisted bilayer graphene (TBG) obtained from \textsc{lammps} simulations. In particular, we consider commensurate structures that have the periodicity of the moir\'e lattice with twist angles defined by \cite{lopes_dos_santos_continuum_2012}
\begin{equation}
    \cos \vartheta_m = \frac{3m^2 + 3m + 1/2}{3m^2 + 3m + 1}.
\end{equation}
Moreover, we place the twist center at the center of a graphene hexagon such that $\vartheta = 0$ corresponds to $AA$ stacking. These structures have point group $D_6 = \left< \mathcal C_{6z}, \mathcal C_{2x} \right>$ where $\mathcal C_{6z}$ is a rotation by $\pi/3$ about the $z$ axis and $\mathcal C_{2x}$ is a $\pi$ rotation about the $x$ axis \cite{zou_band_2018}, as illustrated in Fig.\ \ref{fig:tbgD6}(a).
We then define the displacement fields as
\begin{equation}
    \bm r_{l\sigma} = \bm r_{l\sigma}^0 + \bm u_{l\sigma}(\bm r_{l\sigma}^0) + h_{l\sigma}(\bm r_{l\sigma}^0) \hat z,
\end{equation}
where $l= 1,2$ is the layer, $\sigma = A, B$ the sublattice, and 
$\bm r_{l\sigma}^0= R[(-1)^{l+1}\vartheta/2] \bm \rho_\sigma$ are the rigid coordinates in the absence of relaxation. Here $\bm \rho_\sigma = n_1 \bm a_1 + n_2 \bm a_2 + \bm \delta_\sigma$ are the atomic positions of monolayer graphene with $n_1,n_2 \in \mathds Z$ and $\bm \delta_\sigma$ the sublattice position in the graphene cell.

\subsection{Displacement fields}

\begin{figure}
    \centering
    \includegraphics[width=\linewidth]{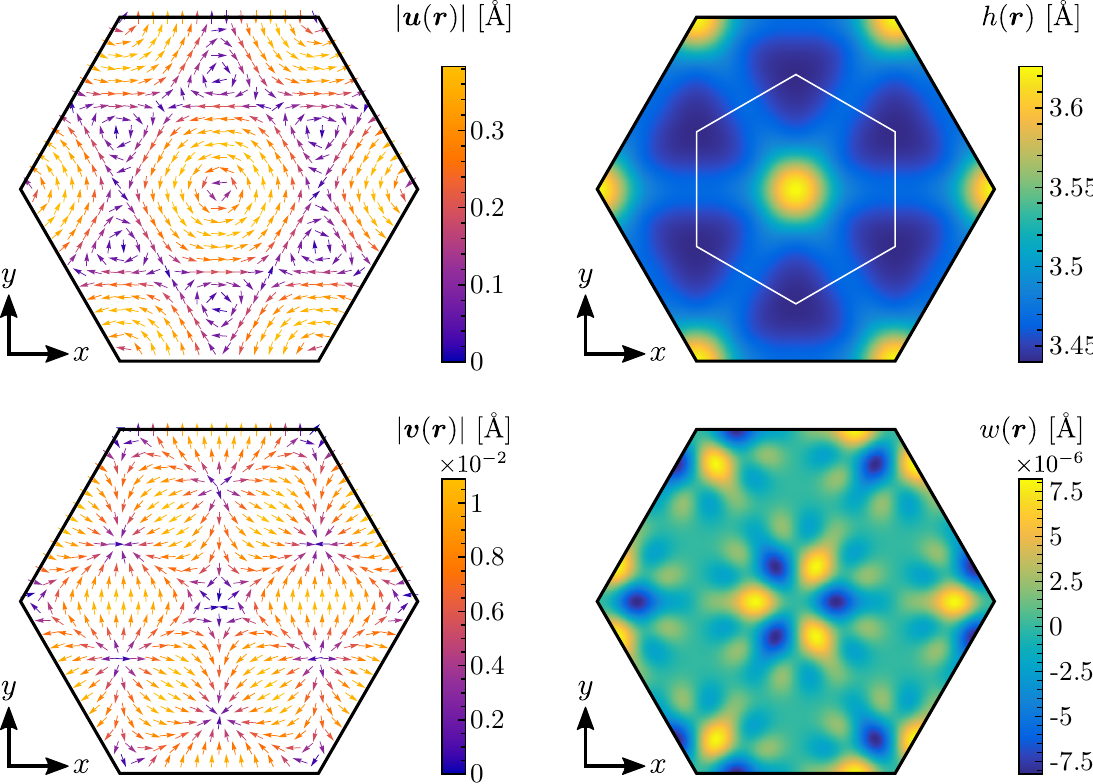}
    \caption{Hetero displacements fields from lattice relaxation for twisted bilayer graphene with $\vartheta \approx 1.018^\circ$ obtained from \textsc{lammps}. The white hexagon gives the moir\'e cell.}
    \label{fig:tbg_displacement}
\end{figure}

Assuming the moir\'e periodicity is preserved after lattice relaxation, we define the smooth fields
\begin{equation}
    \bm u_{l\sigma}(\bm r) = \sum_{\bm g} \bm u_{l\sigma,\bm g} e^{i \bm g \cdot \bm r},
\end{equation}
and similar for out-of-plane displacements. Here $\bm g$ are moir\'e reciprocal vectors and $\bm u_{\bm g} = \bm u_{-\bm g}^*$ are complex Fourier components. In practice, the Fourier components are obtained by taking a discrete Fourier transform of the \textsc{lammps} data. We can now define the acoustic and optical displacement fields for each layer,
\begin{alignat}{2}
    \bm u_l & = ( \bm u_{lA} + \bm u_{lB} ) / 2, \qquad && h_l = ( h_{lA} + h_{lB} ) / 2, \\
    \bm v_l & = \bm u_{lA} - \bm u_{lB}, \qquad && w_l = h_{lA} - h_{lB}.
\end{alignat}
One can now make similar superpositions between layers in terms of homo and hetero displacements. For example, one distinguishes between out-of-plane buckling (homo) and breathing (hetero) displacements. The hetero displacements are given by
\begin{alignat}{2}
    \bm u & = \bm u_1 - \bm u_2, \qquad && h = h_1 - h_2, \\
    \bm v & = \bm v_1 - \bm v_2, \qquad && w = w_1 - w_2,
\end{alignat}
which are shown in Fig.\ \ref{fig:tbg_displacement} for $\vartheta \approx 1.018^\circ$. We see that $\bm u$ is mostly solenoidal, i.e., $\nabla \cdot \bm u \approx 0$. It gives rise to local co-twisting near the AA stacking center (origin) and counter-twist near AB and BA stacking centers \cite{ezzi_analytical_2024}. This reduces the size of AA regions and increases the size of AB and BA regions. Similarly, the acoustic out-of-plane hetero displacements, i.e., the interlayer distance, conforms to the in-plane stacking. On the other hand, the in-plane optical displacement field $\bm v$ has significant volumetric contributions and is over one order of magnitude smaller than $\bm u$, while the out-of-plane field $w$ is about six orders of magnitude smaller than $h$ and can be safely neglected.

\begin{figure}
    \centering
    \includegraphics[width=\linewidth]{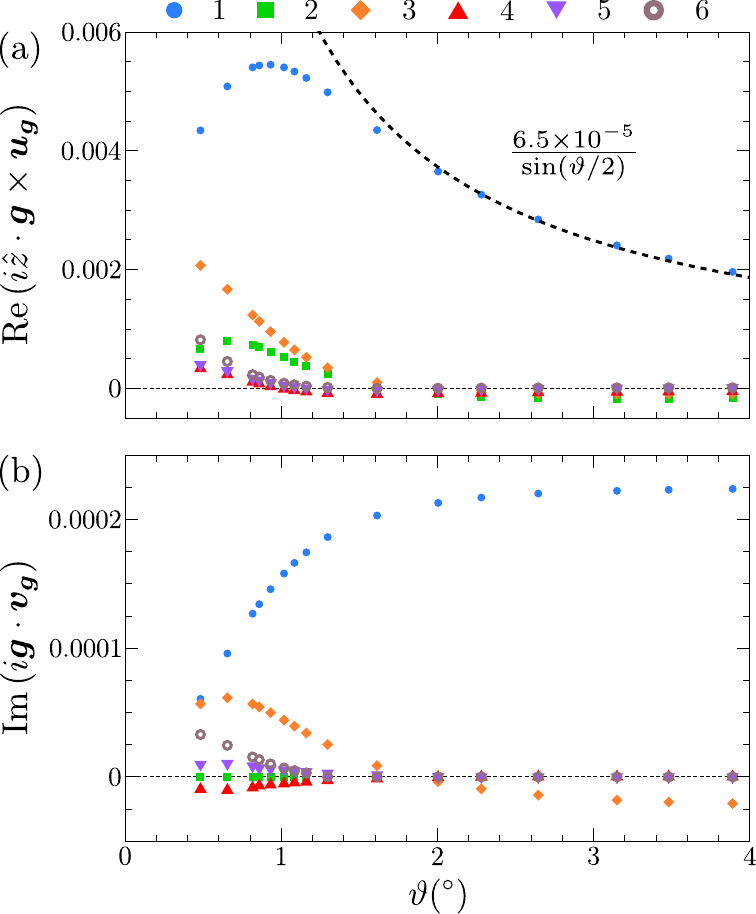}
    \caption{(a) Real part of in-plane transverse Fourier components of the acoustic hetero displacement field for twisted bilayer graphene as a function of twist angle. (b) Imaginary part of in-plane longitudinal Fourier components of the optical hetero displacement field as a function of twist angle. Here the labels correspond to the reciprocal star. Calculated from \textsc{lammps} using a discrete Fourier transform.}
    \label{fig:tbg_fourier}
\end{figure}

The in-plane components can be written using a Helmholtz decomposition. For example,
\begin{equation} \label{eq:ug}
    \bm u_{\bm g} = \frac{u_{\bm g}^\parallel \bm g + u_{\bm g}^\perp \hat z \times \bm g}{ig^2},
\end{equation}
for $g = |\bm g| \neq 0$ and where $u_{\bm g}^\parallel = ( u_{-\bm g}^\parallel)^*$  and $u_{\bm g}^\perp = ( u_{-\bm g}^\perp )^*$ are complex numbers. Here we set $\bm u_{\bm 0} = \bm 0$ since this corresponds to a constant relative shift between layers which does not affect the long-wavelength physics for small twists. These coefficients are related to the divergence and curl:
\begin{align}
    \nabla \times \bm u & = \sum_{\bm g} i\bm g \times \bm u_{\bm g} e^{i \bm g \cdot \bm r} = \hat z \sum_{\bm g} u_{\bm g}^\perp e^{i\bm g \cdot \bm r}, \\
    \nabla \cdot \bm u & = \sum_{\bm g} i\bm g \cdot \bm u_{\bm g} e^{i \bm g \cdot \bm r} = \sum_{\bm g} u_{\bm g}^\parallel e^{i\bm g \cdot \bm r},
\end{align}
which are the rotational and in-plane volumetric components of the displacement gradient. From the \textsc{lammps} simulations, we find that $\bm u$ is dominated by real rotational coefficients while $\bm v$ is mostly given in terms of imaginary volumetric coefficients. In Fig.\ \ref{fig:tbg_fourier}, we show these Fourier coefficients as function of twist angle for the first six reciprocal stars. The scaling of the first Fourier component of the acoustic displacement field yields an estimate of the $V_1 / \mu$ where $V_1$ is the first Fourier coefficient of the adhesion energy \cite{ezzi_analytical_2024,ceferino_pseudomagnetic_2024} and $\mu$ is the shear Lam\'e constant. From this, we estimate that relaxation in our molecular dynamics simulations is about $1.5$ stronger as compared to density-functional theory calculations using the local-stacking approximation \cite{carr_relaxation_2018}. We further find that the $\text{Re} ( u_{\bm g}^\perp )$ [$\text{Im} ( v_{\bm g}^\parallel )$] can be fitted to a polynomial odd [even] in $1/\theta$.

All results obtained from \textsc{lammps} molecular dynamics simulations are consistent with the emergent $D_6$ symmetry of twisted bilayer graphene. To illustrate how symmetries constrain the displacement fields, consider an in-plane symmetry $\mathcal S$. The displacement fields then satisfy $\mathcal S \bm u(\bm r) = \bm u(\mathcal S\bm r)$ and $h(\bm r) = h(\mathcal S \bm r)$. In reciprocal space, one then finds that $u_{\bm g}^\perp$ and $u_{\bm g}^\parallel$ transform as a pseudoscalar and scalar, respectively. Explicitly,
\begin{align}
    h_{\mathcal S\bm g} & = h_{\bm g}, \\
    u_{\mathcal S\bm g}^\parallel & = u_{\bm g}^\parallel, \\
    u_{\mathcal S\bm g}^\perp & = \det(\mathcal S) u_{\bm g}^\perp.
\end{align}
The optical displacements transform similarly except for the fact that any transformation that interchanges the sublattices gives an extra minus sign. For example, in the presence of $\mathcal C_{2z}$ rotation symmetry we have $\bm u(-\bm r) = -\bm u(\bm r)$ and $h(-\bm r) = h(\bm r)$, while $\bm v(-\bm r) = \bm v(\bm r)$ and $w(-\bm r) = -w(\bm r)$. The symmetry-allowed coefficients for the first five reciprocal stars are listed in Table \ref{tab:D6}. Here each reciprocal star consists of six reciprocal vectors closed under $\mathcal C_{6z}$ rotations.
\begin{table}
    \centering
    \begin{tabular}{c | c | c | c | c}
        \Xhline{1pt}
        $m$ & $g/g_1$ & $u_{\bm g}^\perp = i \hat z \times \bm g \cdot \bm u_{\bm g}$ & $u_{\bm g}^\parallel = i \bm g \cdot \bm u_{\bm g}$ & $h_{\bm g}$ \\
        \hline
        $1$ & $1$ & $\mathds R$ & $0$ & $\mathds R$ \\ 
        $2$ & $\sqrt{3}$ & $\mathds R$ & $0$ & $\mathds R$ \\
        $3$ & $2$ & $\mathds R$ & $0$ & $\mathds R$ \\
        $4$ & $\sqrt{7}$ & $\mathds R$ & $\mathds R$ & $\mathds R$ \\
        $5$ & $\sqrt{7}$ & $u_4^\perp$ & $-u_4^\parallel$ & $h_4$\\
        \Xhline{1pt}
    \end{tabular}\\
    \vspace{3mm}
    \begin{tabular}{c | c | c | c | c}
        \Xhline{1pt}
        $m$ & $g/g_1$ & $v_{\bm g}^\perp = i \hat z \times \bm g \cdot \bm v_{\bm g}$ & $v_{\bm g}^\parallel = i \bm g \cdot \bm v_{\bm g}$ & $w_{\bm g}$ \\
        \hline
        $1$ & $1$ & $0$ & $i\mathds R$ & $0$ \\ 
        $2$ & $\sqrt{3}$ & $i\mathds R$ & 0 & $i\mathds R$ \\
        $3$ & $2$ & $0$ & $i\mathds R$ & $0
        $ \\
        $4$ & $\sqrt{7}$ & $i\mathds R$ & $i\mathds R$ & $i\mathds R$ \\
        $5$ & $\sqrt{7}$ & $v_4^\perp$ & $-v_4^\parallel$ & $w_4$\\
        \Xhline{1pt}
    \end{tabular}
    \caption{Symmetry-allowed values of the Fourier coefficients of the hetero displacement fields in the presence of $D_6$ symmetry for the first five reciprocal stars.}
    \label{tab:D6}
\end{table}

\subsection{Pseudomagnetic fields}

\begin{figure*}
    \centering
    \includegraphics[width=\linewidth]{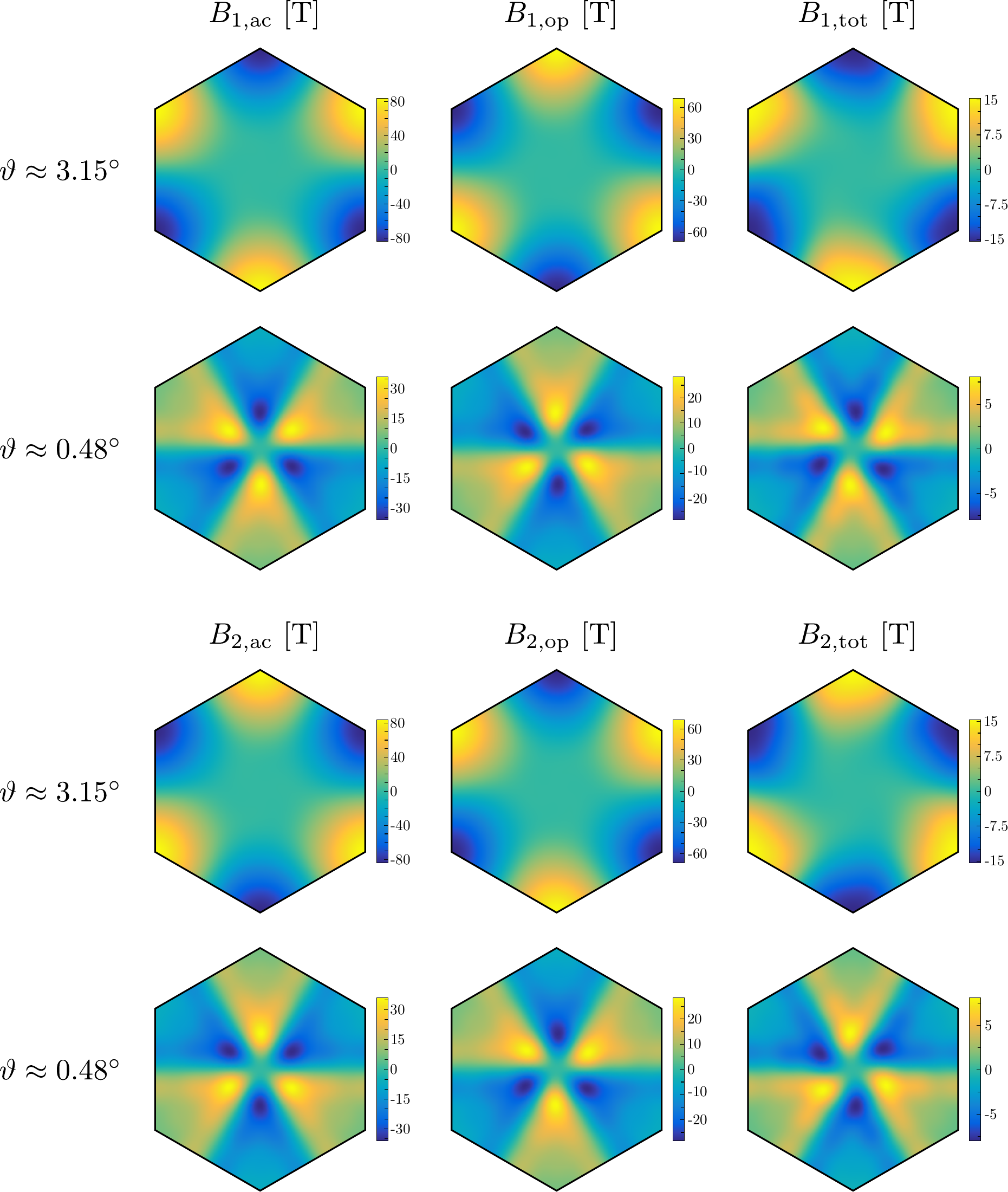}
    \caption{pseudomagnetic fields calculated for twisted bilayer graphene for layers $l=1,2$ ($\pm \vartheta/2$) for two twist angles as indicated. Shown in the moir\'e cell. We show the PMFs from acoustic, optical, and total displacements.}
    \label{fig:tbg_pmf}
\end{figure*}

The valley-preserving symmetries of the emergent moir\'e lattice in small-angle twisted bilayer graphene form the dichromatic group $\text{6\textquotesingle2\textquotesingle2} = \left< \mathcal C_{6z} \mathcal T, \mathcal C_{2y} \right>$ also denoted as $D_6(D_3) = D_3 + (D_6 \setminus D_3 ) \mathcal T$ \cite{dresselhaus_group_2007} where $\mathcal T$ is spinless time-reversal symmetry with $\mathcal T^2 = 1$. These symmetries yield the following constraints on the pseudogauge fields,
\begin{align}
    \bm A_{1,2}(\bm r) & = \bm A_{1,2}(-\bm r) + \nabla \chi, \\
    \bm A_{1,2}(\bm r) & = \mathcal C_{3z}^{-1} \bm A_{1,2}(\mathcal C_{3z} \bm r) + \nabla \chi, \\
    \bm A_1(x,y) & = \begin{pmatrix} -1 & 0 \\ 0 & 1 \end{pmatrix} \bm A_2(-x,y),
\end{align}
where the subscript is the layer index and $\chi$ is a scalar function. This implies that the pseudomagnetic field (PMF) satisfies
\begin{align}
    B_{1,2}(\bm r) & = -B_{1,2}(-\bm r) = B_{1,2}(\mathcal C_{3z}\bm r), \\
    B_1(x,y) & = -B_2(-x,y) = B_2(x,-y), \label{eq:symB}
\end{align}
such that we only need to consider one layer. To verify these symmetries, we show the PMFs for both layers obtained from the acoustic and optical displacement fields in Fig.\ \ref{fig:tbg_pmf} for two twist angles.

\begin{figure}
    \centering
    \includegraphics[width=\linewidth]{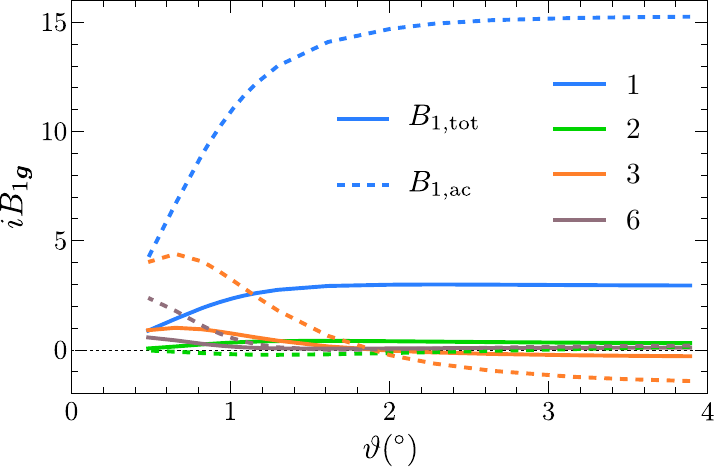}
    \caption{Four largest Fourier coefficients of the pseudomagnetic field in twisted bilayer graphene as a function of twist angle. Shown here for the layer rotated by $+\vartheta/2$. We show both the acoustic (dashed) and total (solid) contributions.}
    \label{fig:tbg_Bg}
\end{figure}

Similar as for the displacement fields, we can write the PMF as a Fourier series
\begin{equation}
    B(\bm r) = \sum_{\bm g \neq \bm 0} B_{\bm g} e^{i\bm g\cdot \bm r},
\end{equation}
where we restrict the sum since the uniform part vanishes for periodic displacement fields, giving a zero net pseudoflux. We show the Fourier components of the acoustic ($B_\text{ac}$) and total PMF ($B_\text{tot} = B_\text{ac} + B_\text{op}$) for the first three reciprocal stars in Fig.\ \ref{fig:tbg_Bg} as a function of twist angle. We see that the PMF is well approximated by
\begin{equation}
    B_{1,2}(x,y) \simeq \pm 2 B_0 \sum_{i=1}^3 \sin(\bm g_i \cdot \bm r), \label{eq:Bapprox}
\end{equation}
where $\bm g_{1,2,3}$ are the shortest nonzero moir\'e reciprocal vectors related by $120^\circ$ rotations. Here $|B_0| \approx 3 \, \text{T}$ for twist angles in the range $1.5^\circ < \vartheta < 4^\circ$ and where we used $\beta = 3.37$. The independence of the magnitude of the PMF for twist angles above the magic angle ($\vartheta \approx 1^\circ$) was also found in other studies that only considered the acoustic part \cite{ezzi_analytical_2024,ceferino_pseudomagnetic_2024}. Note that the form in Eq.\ \eqref{eq:Bapprox} satisfies the symmetry constraints in Eq.\ \eqref{eq:symB} since the first star is symmetric under $x \mapsto -x$ in the presence of $\mathcal C_{3z}$ symmetry. 

\begin{figure}
    \centering
    \includegraphics[width=\linewidth]{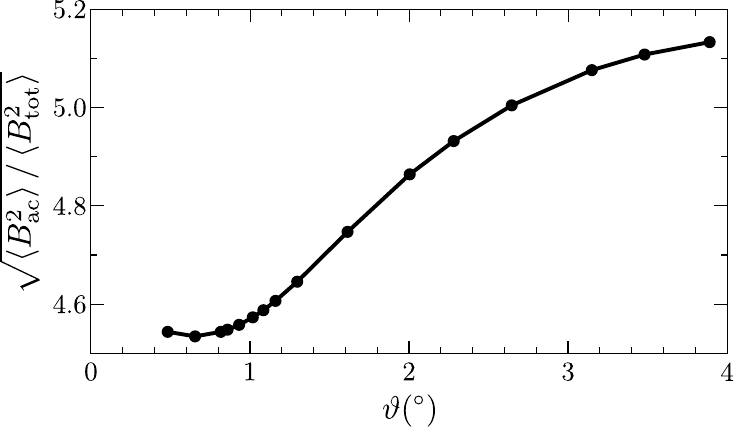}
    \caption{RMS ratio between the acoustic and total PMF in twisted bilayer graphene as a function of twist angle.}
    \label{fig:tbg_Bratio}
\end{figure}

Finally, the root mean square is then given by
\begin{equation}
    \text{RMS} = \sqrt{\left< B^2 \right>} = \sqrt{\sum_{\bm g} |B_{\bm g}|^2},
\end{equation}
which is shown in Fig.\ \ref{fig:fig2} of the main text. If we define the magnetic length from the RMS, one finds $\ell_\text{RMS} > 9 \, \text{nm}$ for the twist angles under consideration. We also show the ratio $\sqrt{\left<B_\text{ac}^2\right> / \left< B_\text{tot}^2 \right>}$ in Fig.\ \ref{fig:tbg_Bratio} as a function of twist angle. This ratio gives an estimate of the reduction factor \cite{woods_electron-phonon_2000,suzuura_phonons_2002} due to optical displacements.

\subsection{Molecular dynamics} 

For the molecular dynamics simulation of twisted bilayer graphene we divide the interactions between interlayer and intralayer. For the interlayer interaction we take the dihedral-angle-corrected registry-dependent potential (DRIP) benchmarked with EXX-RPA DFT calculations \cite{leconte_relaxation_2022}, which has proven to be an improvement in comparison to the usual Kolgomorov-Crespi \cite{kolmogorov_registry-dependent_2005} potential. For the intralayer interactions we use the usual REBO potential \cite{stuart_reactive_2000} with a 2 \r{A} cutoff to avoid interactions between atoms in different layers. For the geometric optimization we enforce periodic boundary conditions and use the "fire" minimization style which uses damped dynamics. This ensures that the system does not get stuck in a local energy minimum. As a stopping criteria, we take the force tolerance to be $10^{-4}$ eV/\r{A} between subsequent steps. 
\begin{figure*}
    \centering
    \includegraphics[width=\linewidth]{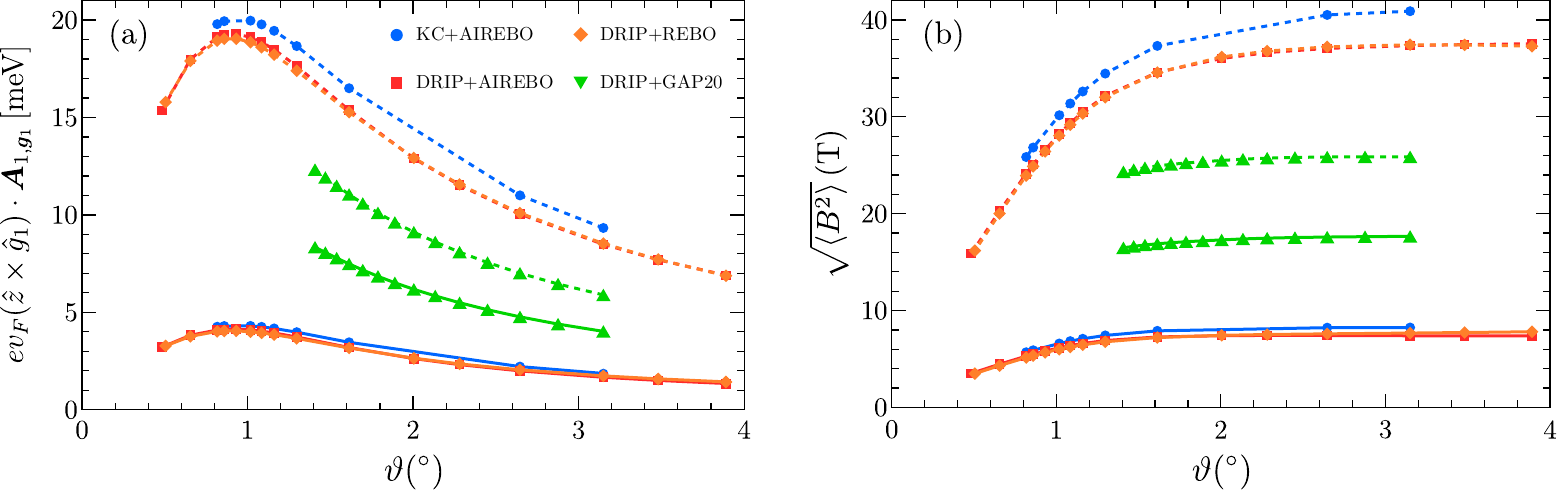}
    \caption{Comparing different MD potentials for pseudogauge fields induced by lattice relaxation in twisted bilayer graphene. (a) Transverse Fourier coefficient of the first star for the effective vector potential as a function of twist angle. (b) Root mean square (RMS) of the PMF as a function of twist angle. In both panels, the solid and dashed lines correspond to the total PMF calculated with Eq.\ \eqref{eq:Adef} of the main text and the acoustic contribution from shear strain, respectively.}
    \label{fig:comparison}
\end{figure*}

\subsection{Comparing different microscopic models}

In Fig.\ \ref{fig:comparison} we show the first-star transverse Fourier component of the vector potential of the first layer, $\hat z \cdot \left( \bm A_{1,\bm g_1} \times \bm g_1 \right)$, and the RMS of the PMF as a function of twist angle for different MD potentials. In the main text we showed results for the interlayer DRIP \cite{leconte_relaxation_2022} and intralayer AIREBO \cite{stuart_reactive_2000} potential. Here we also show results using the Kolgomorov-Crespi (KC) \cite{kolmogorov_registry-dependent_2005} interlayer potential instead of DRIP. Moreover, we also show data \cite{ezzi_analytical_2024,pallewela_private_nodate} for the intralayer REBO \cite{brenner_second-generation_2002} and the same interlayer DRIP as before. We finally also show data for DRIP with the intralayer machine-learning potential GAP20 \cite{rowe_accurate_2020}.

We see that the specific microscopic details, corresponding here to different MD potentials, not only yield different total PMFs, but also different acoustic and optical contributions. Most strikingly, the results for DRIP with GAP20 yield a noticeably smaller reduction as compared to the other cases which have similar reduction factors. Even though DRIP with GAP20 gives smaller acoustic contributions, i.e., less strain compared to using REBO or AIREBO for intralayer interactions, the optical contributions are much smaller such that in the end the total PMF is larger for GAP20.

\section{Corrugated graphene} \label{app:corrugated}

In this section, we consider the structural and electronic properties of monolayer graphene subjected to a periodic corrugation. Specifically, we consider a triangular height modulation with $C_{3v}$ symmetry, commensurate with the graphene lattice and defined by \cite{de_beule_network_2023,mahmud_topological_2023,de_beule_roses_2023}
\begin{equation} \label{eq:hC3v}
    h_\text{sub}(\bm r) = h_0 \sum_{n=1}^3 \cos \left( \G_n \cdot \bm r + \theta \right),
\end{equation}
with amplitude $h_0$ and where $\theta$ controls the shape. The superlattice is defined by the reciprocal vectors
\begin{equation}
    \G_1 = \frac{4\pi}{\sqrt{3}L} \begin{pmatrix} 0 \\ 1 \end{pmatrix}, \qquad \G_{2,3} = \frac{4\pi}{\sqrt{3}L} \begin{pmatrix} \mp \sqrt{3}/2 \\ -1/2 \end{pmatrix},
\end{equation}
and $\G_3 = - \G_1 - \G_2$ where $L = Na$ is the lattice constant of the height modulation. The corresponding lattice vectors $\bm l_{1,2}$ are chosen such that $\bm g_i \cdot \bm l_j = 2 \pi \delta_{ij}$. Here we have taken the coordinate system shown in Fig.\ \ref{fig:graphene_app} with the zigzag direction along the $x$ axis.

The height profile from Eq.\ \eqref{eq:hC3v} preserves $C_{3v} = \left< \mathcal C_{3z}, \mathcal M_x \right>$ symmetry on the superlattice scale where $\mathcal C_{3z}$ is a rotation by $120^\circ$ about the $z$ axis and $\mathcal M_x$ is a mirror $x \mapsto -x$. Note that these are not the microscopic symmetries of the graphene lattice, which are broken by the corrugation. In general, the corrugation breaks $\mathcal C_{2z}$ rotation symmetry since this operation is equivalent to $\theta \mapsto -\theta$. Moreover, 
in the long-wavelength limit $L \gg a$ we can restrict to $\theta \in [0, \pi/3[$. This follows from the fact that $\pm \theta$ are $\mathcal C_{2z}$ partners, while $\theta \mapsto \theta + 2\pi/3$ is equivalent to a translation $y \mapsto y + 2L/\sqrt{3}$. Hence, for the special case $\theta = 0 \,\, \text{mod} \,\, \pi/3$, the point group of the superlattice becomes $C_{6v} = \left< \mathcal C_{6z}, \mathcal M_x \right>$.

\subsection{Symmetry constraints}

To minimize the elastic energy, the corrugated graphene lattice will relax, giving rise to in-plane displacement fields. Since the corrugation is smooth and periodic on the atomic scale, the in-plane and out-of-plane displacement fields can be written in terms of Fourier series,
\begin{align}
     \bm u_\sigma(\bm r) & = \sum_{\G} \bm u_{\sigma,\G} e^{i\G \cdot \bm r}, \\
     h_\sigma(\bm r) & = \sum_{\G} h_{\sigma,\G} e^{i\G \cdot \bm r},
\end{align}
respectively, where $\sigma = A, B$ is the sublattice index, $\G$ are reciprocal lattice vectors of the corrugation, and $h_{\G} = h_{-\G}^*$ and $\bm u_{\G} = \bm u_{-\G}^*$ are Fourier components. Here the uniform components ($|\bm g|=0$) are set to zero as these only result in an overall translation of the graphene.
\begin{figure}
    \centering
    \includegraphics[width=\linewidth]{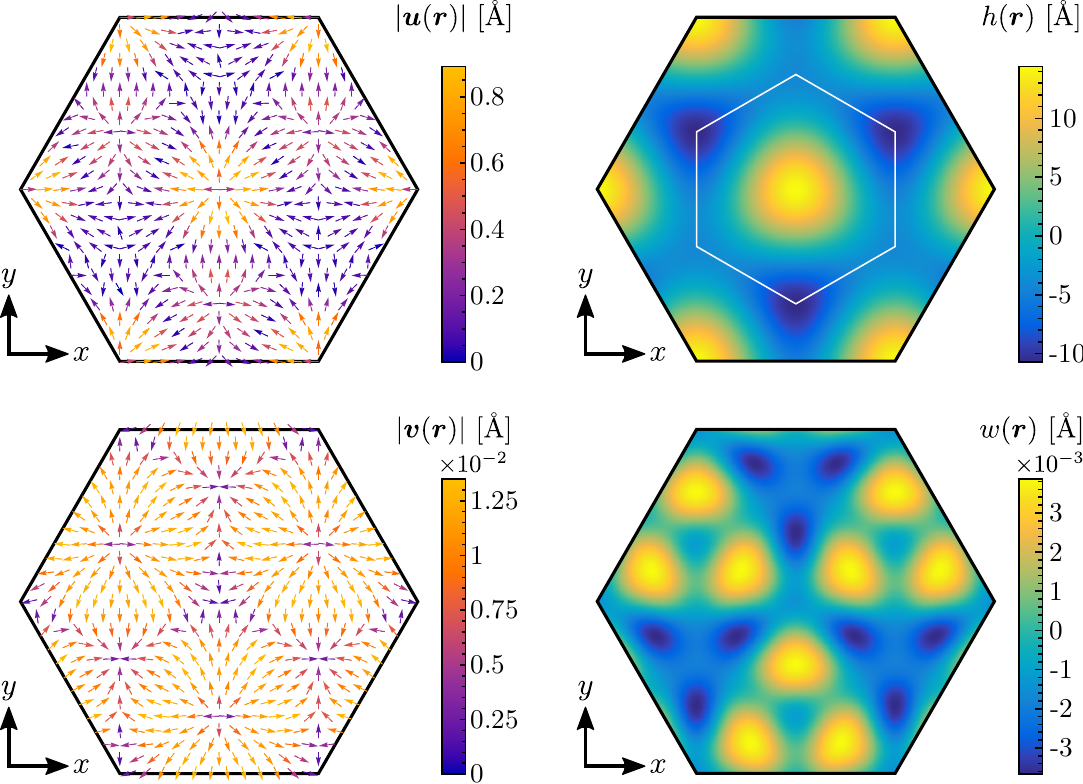}
    \caption{Displacements fields from lattice relaxation for corrugated graphene with $h_0 = 5 \, \text{\r A}$ and $\theta = 15^\circ$ obtained from \textsc{lammps}. The white hexagon gives the supercell.}
    \label{fig:corrugated_displacement}
\end{figure}
\begin{table}
    \begin{tabular}{c | c | c | c | c}
        \Xhline{1pt}
        $m$ & $g/g_1$ & $u_{\bm g}^\perp$ or $v_{\bm g}^\perp$ & $u_{\bm g}^\parallel$ or $v_{\bm g}^\parallel$ & $h_{\bm g}$ or $w_{\bm g}$ \\
        \hline
        $1$ & $1$ & $0$ & $\mathds C$ & $\mathds C$ \\ 
        $2$ & $\sqrt{3}$ & $i\mathds R$ & $\mathds R$ & $\mathds R$ \\
        $3$ & $2$ & $0$ & $\mathds C$ & $\mathds C$ \\
        $4$ & $\sqrt{7}$ & $\mathds C$ & $\mathds C$ & $\mathds C$ \\
        $5$ & $\sqrt{7}$ & $-u_4^\perp$ & $u_4^\parallel$ & $h_4$\\
        \Xhline{1pt}
    \end{tabular}
    \caption{Fourier coefficients for the first five stars consistent with $C_{3v} = \left< \mathcal C_{3z}, \mathcal M_x \right>$ symmetry. The fourth and fifth star are degenerate and related by $\mathcal M_x$ ($x \mapsto -x$). Since no symmetries exchange sublattices, the constraints are the same for acoustic and optical fields. However, for $\theta = 0 \, \text{mod} \, \pi/3$ the point group is $C_{6v} = \left< \mathcal C_{6z}, \mathcal M_x \right>$. In this case, $\mathcal C_{2z}$ requires that all nonzero coefficients are also real for $\bm u(\bm r)$ and $h(\bm r)$, and imaginary for $\bm v(\bm r)$ and $w(\bm r)$.}
    \label{tab:ug}
\end{table}
\begin{figure}
    \centering
    \includegraphics[width=\linewidth]{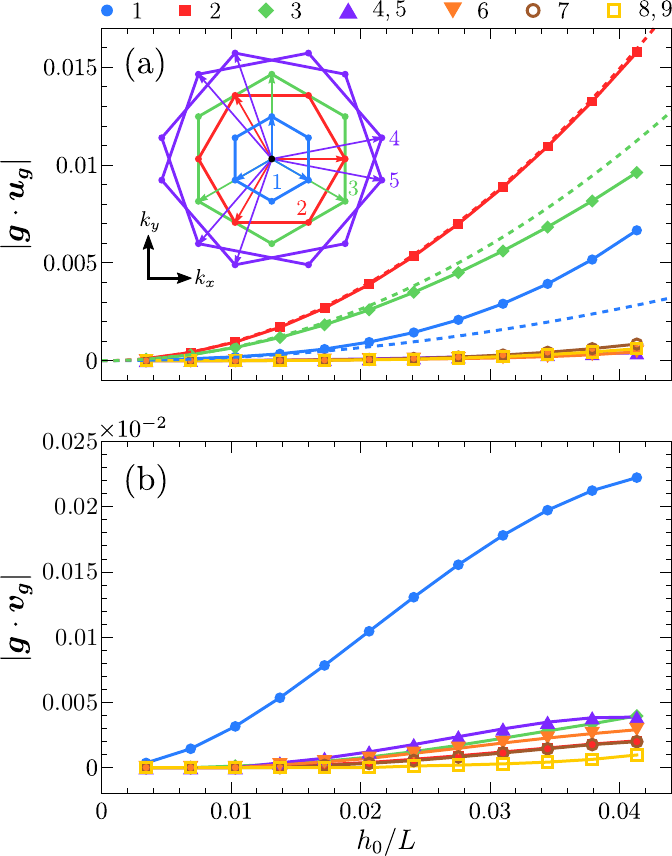}
    \caption{Volumetric Fourier components of the in-plane (a) acoustic and (b) optical displacements fields for corrugated graphene with $\theta = 15^\circ$ obtained from \textsc{lammps} as a function of $h_0/L$ for the first nine stars. In (a) the dashed lines give the results from continuum elasticity and the inset shows the first five reciprocal stars in momentum space.}
    \label{fig:corrugated_fourier}
\end{figure}

As described in the main text, we can then define acoustic and optical displacement fields. The displacement fields calculated with \textsc{lammps} for a corrugated substrate with $h_0 = 5 \, \text{\r A}$ and $\theta = 15^\circ$ are shown in Fig.\ \ref{fig:corrugated_displacement}. Since the relaxed structure is assumed to be adiabatically connected to the rigid corrugation, the displacements fields obey the same symmetries. In particular, due to $\mathcal C_{3z}$ symmetry and the reality of the fields, the displacements are characterized by three complex numbers $u_m^\parallel$, $u_m^\perp$, and $h_m$ for each star of reciprocal vectors, which is indexed by $m$. Here, we define a reciprocal star by six reciprocal vectors that are related by $\mathcal C_{6z}$ and we use a Helmholtz decomposition for the in-plane fields. 

We show the symmetry-allowed values for a corrugation with $C_{3v}$ symmetry in Table \ref{tab:ug} for the first five stars. We also show the volumetric Fourier coefficients of the in-plane displacements, i.e., projected on the $xy$ plane, in Fig.\ \ref{fig:corrugated_fourier} for $\theta = 15^\circ$ as a function of $h_0/L$ for the first nine reciprocal stars. 

\subsection{Continuum elasticity}

The long-wavelength acoustic displacements in graphene can be modeled using the continuum theory of elasticity \cite{landau_theory_1986}. Here one views the graphene as an elastically isotropic membrane. The elastic potential energy \cite{nelson_david_statistical_2004} and substrate interaction are modeled as \cite{guinea_gauge_2008}
\begin{align}
    H_\text{elas} & = \frac{1}{2} \int d^2 \bm r \left[ \lambda u_{ii} u_{ii} + 2 \mu u_{ij} u_{ji} + \kappa \left( \nabla^2 h \right)^2 \right], \label{eq:Helas} \\
    H_\text{sub} & = \frac{\gamma}{2} \int d^2 \bm r \left[ h(\bm r) - h_\text{sub}(\bm r) \right]^2, \label{eq:Hsub} 
\end{align}
where $\lambda$ and $\mu$ are in-plane Lam\'e constants, $\kappa$ is the out-of-plane bending rigidity, and $\gamma$ controls the interaction with the substrate. Here summation over repeated indices is implied and
\begin{equation} \label{eq:straintensor}
    u_{ij}(\bm r) = \frac{1}{2} \left( \frac{\partial u_j}{\partial r_i} + \frac{\partial u_i}{\partial r_j} + \frac{\partial h}{\partial r_i} \frac{\partial h}{\partial r_j} \right),
\end{equation}
is the strain tensor with $i,j = x,y$. We further assume that the graphene is pinned to the substrate such that $h(\bm r) = h_\text{sub}(\bm r)$. This is justified in the limit $L \gg \left( \kappa / \gamma \right)^{1/4} \approx 1 \; \text{nm}$ \cite{guinea_gauge_2008} where $L$ is periodicity of the corrugation and the numerical value is for graphene on SiO$_2$ \cite{sabio_electrostatic_2008}. In this case, the interaction with the substrate dominates over the curvature term [last term of Eq.\ \eqref{eq:Helas}] since the latter scales as $\kappa h^2 / L^4$ while the former scales as $\gamma h^2$. In this limit, $H_\text{elas}$ is a functional of the in-plane field only.

Under these assumptions, the elastic energy density
\begin{widetext}
\begin{equation}
    \mathcal H_\text{elas} = \frac{1}{A_c} \int_\text{cell} d^2 \bm r \left[ \left( \frac{\lambda}{2} + \mu \right) \left( u_{xx}^2 + u_{yy}^2 \right) + \lambda u_{xx} u_{yy} + 2 \mu u_{xy}^2 \right] + \text{constant},
\end{equation}
\end{widetext}
where $A_c$ is the area of the supercell defined by the periodic height modulation. For the triangular height profile, we have $A_c = \sqrt{3} L^2 / 2$. For convenience, we define the tensor
\begin{equation}
        f_{ij}(\bm r) \equiv \left[ \partial_i h(\bm r) \right] \left[ \partial_j h(\bm r) \right] = \sum_{\G} f_{ij\G} \, e^{i \G \cdot \bm r},
\end{equation}
where
\begin{equation}
    f_{ij\G} = - \sum_{\G'} h_{\G'} h_{\G - {\G}'} g_i' \left( g_j - g_j'\right),
\end{equation}
with $f_{ij,-\G} = f_{ij\G}^*$. The strain tensor thus becomes
\begin{equation}
    u_{ij}(\bm r) = \frac{1}{2} \sum_{\G} \left[ i \left( g_i u_{j\G} + g_j u_{i\G} \right) + f_{ij\G} \right] e^{i \G \cdot \bm r},
\end{equation}
where we set $u_{i\bm 0} = 0$ since this amounts to a uniform translation. In terms of the Helmoltz decomposition, 
\begin{align}
    u_{xx\bm g} & = \frac{g_x \left( g_x u_{\bm g}^\parallel - g_y u_{\bm g}^\perp \right)}{g^2} + f_{xx\bm g}, \\
    u_{yy\bm g} & = \frac{g_y \left( g_y u_{\bm g}^\parallel + g_x u_{\bm g}^\perp \right)}{g^2} + f_{yy\bm g}, \\
    u_{xy\bm g} & = \frac{2 g_x g_y u_{\bm g}^\parallel + \left( g_x^2 - g_y^2 \right) u_{\bm g}^\perp}{g^2} + f_{xy\bm g},
\end{align}
with $g = |\bm g|$ and where $u_{\bm g}^\perp$ and $u_{\bm g}^\parallel$ are the rotational and volumetric components of the in-plane displacement field, respectively. Plugging the Fourier expansions into the energy density $\mathcal H_{\text{elas}}$ gives \cite{phong_boundary_2022}
\begin{widetext}
\begin{align}
     \frac{1}{A_c} \int_\text{cell} d^2 \bm r \, u_{ii}^2 & = \frac{1}{A_c} \sum_{\G,\G'} \int d^2 \bm r \left( i g_i u_{i\G} + \frac{f_{ii \G}}{2} \right) \left( i g_i' u_{i\G'} + \frac{f_{ii \G'}}{2} \right) e^{i ( \G + \G' ) \cdot \bm r} \\
     & = \sum_{\G} \left| i g_i u_{i\G} + \frac{f_{ii \G}}{2} \right|^2, \\
     \frac{1}{A_c} \int_\text{cell} d^2 \bm r \, u_{xx} u_{yy} & = \sum_{\G} \left( i g_x u_{x\G} + \frac{f_{xx \G}}{2} \right) \left( -i g_y u_{y\G}^* + \frac{f_{yy \G}^*}{2} \right) \\
     & = \frac{1}{2} \sum_{\G} \left[ \left( i g_x u_{x\G} + \frac{f_{xx \G}}{2} \right) \left( -i g_y u_{y\G}^* + \frac{f_{yy \G}^*}{2} \right) + \text{c.c.} \right], \\
     \frac{1}{A_c} \int_\text{cell} d^2 \bm r \, u_{(xy)}^2 & = \frac{1}{4} \sum_{\G} \left( i g_x u_{y\G} + i g_y u_{x\G} + f_{xy\G} \right) \left( -i g_x u_{y\G}^* - i g_y u_{x\G}^* + f_{xy\G}^* \right).
\end{align}
We obtain
\begin{align}
    \mathcal H_\text{elas} & = \left( \frac{\lambda}{2} + \mu \right) \sum_{\G} \left( i g_x u_{x\G} + \frac{f_{xx \G}}{2} \right) \left( -i g_x u_{x\G}^* + \frac{f_{xx \G}^*}{2} \right) \\
    & + \left( \frac{\lambda}{2} + \mu \right) \sum_{\G} \left( i g_y u_{y\G} + \frac{f_{yy \G}}{2} \right) \left( -i g_y u_{y\G}^* + \frac{f_{yy \G}^*}{2} \right) \\
    & + \frac{\lambda}{2} \sum_{\G} \left[ \left( i g_x u_{x\G} + \frac{f_{xx\G}}{2} \right) \left( -i g_y u_{y\G}^* + \frac{f_{yy\G}^*}{2} \right) + \text{c.c.} \right] \\
    & + \frac{\mu}{2} \sum_{\G} \left( i g_x u_{y\G} + i g_y u_{x\G} + f_{xy\G} \right) \left( -i g_x u_{y\G}^* - i g_y u_{x\G}^* + f_{xy\G}^* \right) + \frac{\kappa}{2} \sum_{\G} | h_{\G} |^2 g^4.
\end{align}
Minimizing with respect to $u_{i\G}^*$ for nonzero $\bm g$ yields the solutions for the Fourier components $u_{i\G}$ in terms of $f_{ij\G}$ (and thus $h_{i\G}$). We find
\begin{align}
    \frac{\partial \mathcal H_\text{elas}}{\partial u_{x\G}^*} & = -i g_x \left[ \left( \lambda + 2\mu \right) \left( i g_x u_{x\G} + \frac{f_{xx \G}}{2} \right) + \lambda \left( i g_y u_{y\G} + \frac{f_{yy \G}}{2} \right) \right] - i \mu g_y \left( i g_x u_{y\G} + i g_y u_{x\G} + f_{xy\G} \right), \\
    \frac{\partial \mathcal H_\text{elas}}{\partial u_{y\G}^*} & = -i  g_y \left[ \left( \lambda + 2 \mu \right) \left( i g_y u_{y\G} + \frac{f_{yy \G}}{2} \right) + \lambda \left( i g_x u_{x\G} + \frac{f_{xx \G}}{2} \right) \right] - i \mu g_x \left( i g_x u_{y\G} + i g_y u_{x\G} + f_{xy\G} \right),
\end{align}
and setting these equations equal to zero, gives
\begin{align}
    u_{x\G} & = \frac{i}{2 \left( \lambda + 2 \mu \right) g^4} \left\{ f_{xx\G} g_x \left[ g_x^2 \left( \lambda + 2 \mu \right) + g_y^2 \left( 3 \lambda + 4 \mu \right) \right] + \left( f_{yy\G} g_x - 2 f_{xy\G} g_y \right) \left[ g_x^2 \lambda - g_y^2 \left( \lambda + 2 \mu \right) \right]  \right\}, \label{eq:uxg} \\
    u_{y\G} & = \frac{i}{2 \left( \lambda + 2 \mu \right) g^4} \left\{ f_{yy\G} g_y \left[ g_y^2 \left( \lambda + 2 \mu \right) + g_x^2 \left( 3 \lambda + 4 \mu \right) \right] + \left( f_{xx\G} g_y - 2 f_{xy\G} g_x \right) \left[ g_y^2 \lambda - g_x^2 \left( \lambda + 2 \mu \right) \right]  \right\}, \label{eq:uyg}
\end{align}
and
\begin{align}
    u_{\bm g}^\perp & = i \bm g \times \bm u_{\bm g} = \frac{\left( f_{xx\bm g} - f_{yy\bm g} \right) g_x g_y + f_{xy\bm g} \left( g_y^2 - g_x^2 \right)}{g^2} = \frac{1}{g^2} \sum_{\bm g'} h_{\bm g'} h_{\bm g-\bm g'} \left( \bm g' \cdot \bm g \right) \left( \hat z \cdot \bm g' \times \bm g \right), \\
    u_{\bm g}^\parallel & = i \bm g \cdot \bm u_{\bm g} = \frac{\mu \mathcal F_{\bm g}}{\lambda + 2 \mu} - \frac{f_{xx\bm g} + f_{yy\bm g}}{2},
\end{align}
with $\mathcal F_{\bm g} = \left( g_x^2 f_{yy\G} - 2 g_x g_y f_{xy\G} + g_y^2 f_{xx\G} \right) / g^2$. We find that $u_{\bm g}^\perp = 0$ when $h(\bm r)$ is restricted to the first star. For completeness, we also give the explicit forms for $f_{ij\bm g}$ for the height profile from Eq.\ \eqref{eq:hC3v}:
\begin{align}
    \bm f_{\bm g_1} & = \frac{2\pi^2}{3} \frac{h_0^2 e^{-2i\theta}}{L^2} \begin{pmatrix} 3 & 0 \\ 0 & -1 \end{pmatrix}, \\
    \bm f_{\bm g_1+2\bm g 2} & = \frac{2\pi^2}{3} \frac{h_0^2}{L^2} \begin{pmatrix} -3 & 0 \\ 0 & 1 \end{pmatrix}, \\
    \bm f_{2\bm g_1} & = \frac{2\pi^2}{3} \frac{h_0^2 e^{2i\theta}}{L^2} \begin{pmatrix} 0 & 0 \\ 0 & -2 \end{pmatrix},
\end{align}
which transform as $\bm f_{\mathcal S\bm g} = \mathcal S \bm f_{\bm g} \mathcal S^{-1}$.
\end{widetext}

Using the relations
\begin{equation}
    \mu = \frac{E}{2 ( 1 + \nu )}, \qquad \lambda = \frac{\nu E}{1 - \nu^2},  
\end{equation}
for isotropic linear elastic two-dimensional materials, with $E$ the Young modulus and $\nu$ the Poisson ratio of graphene, we obtain \cite{guinea_gauge_2008,phong_boundary_2022}
\begin{align}
    u_{xx\G} + u_{yy\G} & = \frac{1-\nu}{2} \mathcal F_{\bm g}, \label{eq:trace} \\
    u_{xx\G} - u_{yy\G} & = \frac{1+\nu}{2} \frac{g_y^2 - g_x^2}{g^2} \mathcal F_{\bm g}, \label{eq:shear1} \\
    u_{xy\G} + u_{yx\G} & = -\frac{1+\nu}{2} \frac{2 g_x g_y}{g^2} \mathcal F_{\bm g}, \label{eq:shear2}
\end{align}
for nonzero $\bm g$ and $u_{ij\bm 0} = f_{ij\bm 0}/2$. These are the volumetric and shear strains, respectively. In this work we use the value $\nu = 0.165$ from experiment \cite{lee_measurement_2008}. Importantly, since $|\mathcal F_{\G}| \sim h^2 / L^2$ we expect the linear theory to be valid only for $L \gg h$.

For the triangular corrugation with $C_{3v}$ symmetry, continuum elasticity gives a relaxed in-plane acoustic displacement field that is irrotational such that $u_{\bm g}^\perp = 0$ even though it is symmetry allowed in some stars. The volumetric components are finite only in the first three stars with
\begin{align}
    u_1^\parallel = u_{\bm g_1}^\parallel & = \left( 1 - 3 \nu \right) \frac{\pi^2}{3} \frac{h_0^2}{L^2} \, e^{-2i\theta}, \\
    u_2^\parallel = u_{\bm g_1+2\bm g_2}^\parallel & = \left( 3 - \nu \right) \frac{\pi^2}{3}\frac{h_0^2}{L^2}, \label{eq:beta2} \\
    u_3^\parallel = u_{2\bm g_1}^\parallel & = \frac{2\pi^2}{3} \frac{h_0^2}{L^2} \, e^{2i\theta},
\end{align}
consistent with the symmetry analysis. Hence, continuum elasticity yields $\bm u = \nabla \phi$ with $\phi_{\bm g} = -u_{\bm g}^\parallel/g^2$.

\subsection{Electronic band structure}

In the presence of a periodic scalar and gauge fields, the valley-projected electronic continuum Hamiltonian from Eq.\ \eqref{eq:Heffapp} can be diagonalized by Fourier transformation,
\begin{equation}
    \psi_\tau(\bm r) = \frac{1}{\sqrt{A}} \sum_{\bm k} \sum_{\G} e^{i (\bm k - \G ) \cdot \bm r} \, c_{\tau,\bm k - \G},
\end{equation}
where $A$ is the system size and the sum over $\bm k$ is restricted to the superlattice Brillouin zone (SBZ) and $c_{\tau,\bm k-\G}^\dag$ ($c_{\tau,\bm k-\G}$) are two-component creation (annihilation) operators that create a fermion in valley $\tau$ with momentum $\bm k - \G$. The Hamiltonian becomes
\begin{widetext}
\begin{equation}
    H = \frac{1}{A} \sum_{\tau} \sum_{\bm k, \bm k'} \sum_{\G, \G'} \int d^2\bm r \, c_{\tau,\bm k'-\G'}^\dag e^{-i(\bm k' - \G') \cdot \bm r} \left\{ \hbar v_F \left[ \bm k - \G + \frac{\tau e}{\hbar} \, \bm A(\bm r) \right] \cdot \left( \tau \sigma_x, \sigma_y \right) + V(\bm r) \sigma_0 \right\} e^{i(\bm k - \G) \cdot \bm r} c_{\tau,\bm k-\G}.
\end{equation}
\end{widetext}
Next, for any function $f(\bm r)$ with the periodicity of the superlattice, we have
\begin{equation}
    \int d^2 \bm r \, e^{-i(\bm k' - \G') \cdot \bm r} \, f(\bm r) \, e^{i(\bm k - \G) \cdot \bm r} = A \delta_{\bm k\bm k'} f_{\G-\G'},
\end{equation}
where we used that $\bm k - \bm g$ with $\bm k$ in the SBZ and $\bm g$ a reciprocal vector of the superlattice, is a unique momentum decomposition. We further used $\int d^2 \bm r = \sum_{\bm R} \int_\text{cell} d^2\bm r$ with $\sum_{\bm R} e^{i \bm k \cdot \bm R} = N \delta_{\bm k\bm 0}$ where the sum runs over superlattice cells. We then obtain
\begin{widetext}
\begin{equation}
    H = \sum_{\tau} \sum_{\bm k} \sum_{\G, \G'} c_{\tau,\bm k-\G'}^\dag \left\{ \hbar v_F \left[ \left( \bm k - \G \right) \delta_{\G\G'} + \frac{\tau e}{\hbar} \, \bm A_{\G-\G'} \right] \cdot \left( \tau \sigma_x, \sigma_y \right) + V_{\G-\G'} \sigma_0 \right\} c_{\tau,\bm k-\G},
\end{equation}  
\end{widetext}
which can be diagonalized numerically by taking a sufficient number of $\G$ vectors for convergence. All results shown in this work were obtained with a cutoff $|\G| < 12k_0$ where $k_0 = 4\pi / 3L$.

For the triangular periodic height profile that is aligned with the zigzag direction, the electronic continuum theory has wallpaper group 14 ($p3m1$) with point group $C_{3v}$ and superlattice translations. The symmetries of the valley-projected Hamiltonian are generated by $\mathcal C_{3z}$ and $\mathcal M_x \mathcal T$ where $\mathcal M_x$ is a mirror symmetry ($x \mapsto -x$) and $\mathcal T$ is spinless time-reversal symmetry with $\mathcal T^2 = 1$. While $\mathcal M_x$ and $\mathcal T$ both interchange valleys, their combination leaves the valleys invariant. This yields the magnetic point group $3m\text{\textquotesingle}$ also denoted as $C_{3v}(C_3)$ \cite{dresselhaus_group_2007}. 

\begin{figure}
    \centering
    \includegraphics[width=\linewidth]{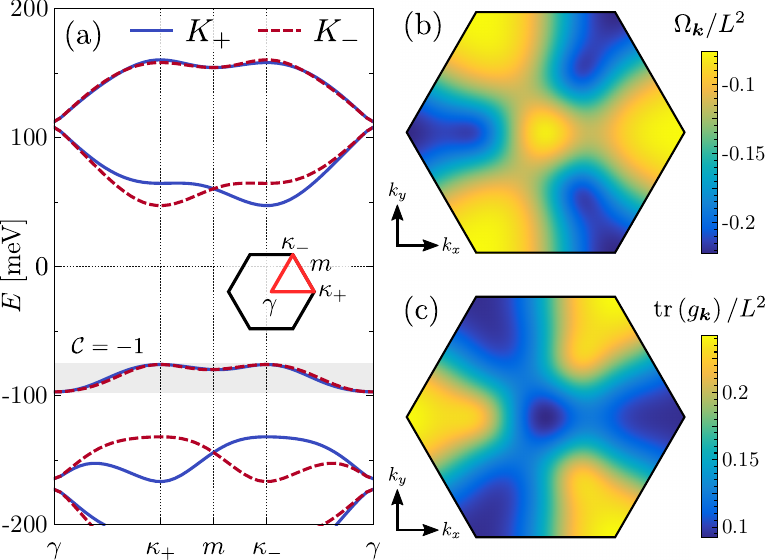}
    \caption{(a) Energy bands calculated with the continuum model for $h_0 = 6 \, \text{\r A}$, $\theta = 15^\circ$, and $V_0 = 72.4 \, \text{meV}$. Solid and dashed lines correspond to valley $K_\pm$, respectively, and the first valence band is highlighted in gray with the valley Chern number indicated. (b) Valley-projected Berry curvature and (c) trace of the quantum metric for the first valence band shown in (a) for valley $K_+$.}
    \label{fig:corrugated_bands_sm}
\end{figure}

In Fig.\ \ref{fig:corrugated_bands_sm}(a) we show the band structure calculated with the continuum model in the presence of a perpendicular electric field $V(\bm r) = V_0 h(\bm r)/h_0$ where $h(\bm r)$ is the relaxed height profile. Parameters are $h_0 = 6 \, \text{\r A}$, $\theta = 15^\circ$, and $V_0 = 71 \, \text{meV}$ corresponding to Fig.\ \ref{fig:fig4}(c) and (d). We also show the Berry curvature and quantum metric of the first valence band in valley $K_+$.

\subsection{Quantum geometry}

We calculate the quantum geometry of an isolated Bloch band in a given valley with the gauge-invariant product. To this end, consider a square plaquette in the Brillouin zone of area $\delta^2$ centered at $\bm k$ with corners: $\bm k_1 = \bm k + \tfrac{\delta}{2}(-1, -1)$, $\bm k_2 = \bm k + \tfrac{\delta}{2}(-1, 1)$, $\bm k_3 = \bm k + \tfrac{\delta}{2}(1, 1)$, and $\bm k_4 = \bm k + \tfrac{\delta}{2}(1, -1)$. The gauge-invariant product is then given by 
\begin{equation}
    \prod_{m=1}^4 \langle u_{\bm k_m} | u_{\bm k_{m+1}} \rangle,
\end{equation}
where $\bm k_5 = \bm k_1$. It is straightfoward to show that
\begin{align}
    \text{tr} \left( g_{\bm k} \right) & = \lim_{\delta \rightarrow 0} \delta^{-2} \, \text{Re} \left( 1 - \prod_{m=1}^4 \langle u_{\bm k_m} | u_{\bm k_{m+1}} \rangle \right), \\
    \Omega_{\bm k} & = \lim_{\delta \rightarrow 0} \delta^{-2} \, \arg \prod_{m=1}^4 \langle u_{\bm k_m} | u_{\bm k_{m+1}} \rangle,
\end{align}
where
\begin{align}
    g_{\bm k}^{\mu\nu} & = \text{Re} \left( \langle \partial^\mu u_{\bm k} | \partial^\nu u_{\bm k} \rangle \right) + \langle u_{\bm k} | \partial^\mu u_{\bm k} \rangle \langle u_{\bm k} | \partial^\nu u_{\bm k} \rangle, \\
    \Omega_{\bm k} & = -2 \, \text{Im} \left( \langle \partial_{k_x} u_{\bm k} | \partial_{k_y} u_{\bm k} \rangle \right),
\end{align}
is the (single band) Fubini-Study quantum metric with $\text{tr} \left( g_{\bm k} \right) = g_{\bm k}^{xx} + g_{\bm k}^{yy}$ and Berry curvature, respectively. They form the real and imaginary components of the (single band) quantum geometric tensor,
\begin{align}
    Q_{\bm k}^{\mu\nu} & = \langle \partial^\mu u_{\bm k} | \left( 1 - | u_{\bm k} \rangle \langle u_{\bm k} | \right) | \partial^\nu u_{\bm k} \rangle \\
    & = g_{\bm k}^{\mu\nu} - \frac{i}{2} \, \epsilon^{\mu\nu} \Omega_{\bm k}.
\end{align}
One can further show that \cite{roy_band_2014}
\begin{equation}
    \text{tr}\left(g_{\bm k}\right) \geq | \Omega_{\bm k} |,
\end{equation}
which is the trace inequality. The trace condition refers to the saturation of this bound and holds for Landau levels. One route of engineering flat bands that may host fractional Chern insulators at fractional filling of the band, is so-called Landau level mimicry \cite{lee_band_2017}. For example, a flat Bloch band with unit Chern number that satisfies the trace condition emulates the lowest Landau level, whose exact ground state at one over odd integer filling in the presence of short-range interactions is given by a Laughlin-type wave function \cite{wang_exact_2021}.

\subsection{Valley polarization}

The tight-binding band structure always features Kramers' pairs of minibands that approximately belong to different valleys since the long-wavelength corrugation does not couple the valleys. We quantify this from the expectation value of the valley polarization $P_v$ \cite{manesco_correlations_2020,manesco_correlation-induced_2021,ramires_electrically_2018,lopez-bezanilla_electrical_2020}. The simplest choice for $P_v$ is given in terms of the next-nearest neighbor Haldane hopping \cite{haldane_model_1988},
\begin{equation}
    P_v = \frac{1}{i3\sqrt{3}} \sum_{\left<\left< i,j \right> \right>} e^{3i\phi_{ij}} c_i^\dag c_j,
\end{equation}
where $\phi_{ij}$ is the angle of the next-nearest neighbor bond vector with the $x$ axis. For pristine graphene, one can diagonalize $P_v$ in momentum space:
\begin{equation}
    P_v = \sum_{\bm k} g(\bm k) c_{\bm k}^\dag c_{\bm k},
\end{equation}
with $g(\bm K_\pm + \bm q) = \pm 1 + \mathcal O(q^2a^2)$. In the presence of potentials that vary slowly with respect to the graphene lattice constant $a$, the valleys remain approximately decoupled such that $\left< P_v \right> \approx \pm 1$ for valley $\bm K_\pm$.

\subsection{Molecular dynamics}

For the \textsc{lammps} simulations, we consider a graphene sheet spanned by $\bm l_{1,2}$ with $L = |\bm l_{1,2} | = 59a \approx 14.5~\text{nm}$. In order to model a generic substrate that induces a smooth van der Waals force on the graphene sheet, we use a honeycomb lattice with lattice constant $a/3$. This ensures commensurability with the graphene and an interaction potential that is smooth on the graphene lattice scale due to the higher density of the substrate. We then apply the out-of-plane displacement field given by Eq.\ \ref{eq:hC3v} to the substrate for $h_0$ ranging from 0.05~nm to 0.6~nm in steps of 0.05~nm and for $\theta = 15^\circ$. Energy minimization was performed with the \textsc{lammps} code \cite{thompson_lammps_2022} using the FIRE minimization procedure, using force tolerance of $10^{-6}$ eV/\r{A} as a stop criteria. To describe the potential energy landscape of the system we use a combination of the AIREBO interatomic potential \cite{stuart_reactive_2000} for short-range interactions between carbon atoms in the graphene sheet with a cutoff of 2 \r{A}, and a 12-6 Lennard-Jones potential for the interaction between the graphene and the substrate. The Lennard-Jones interaction was parametrized with an energy constant of 10 meV, a zero-crossing distance of 3.7 \r{A}, and a cut-off radius of 10 \r{A}. 

\section{Graphene flake subjected to triaxial stress}

\begin{figure*}
    \centering
    \includegraphics[width=\linewidth]{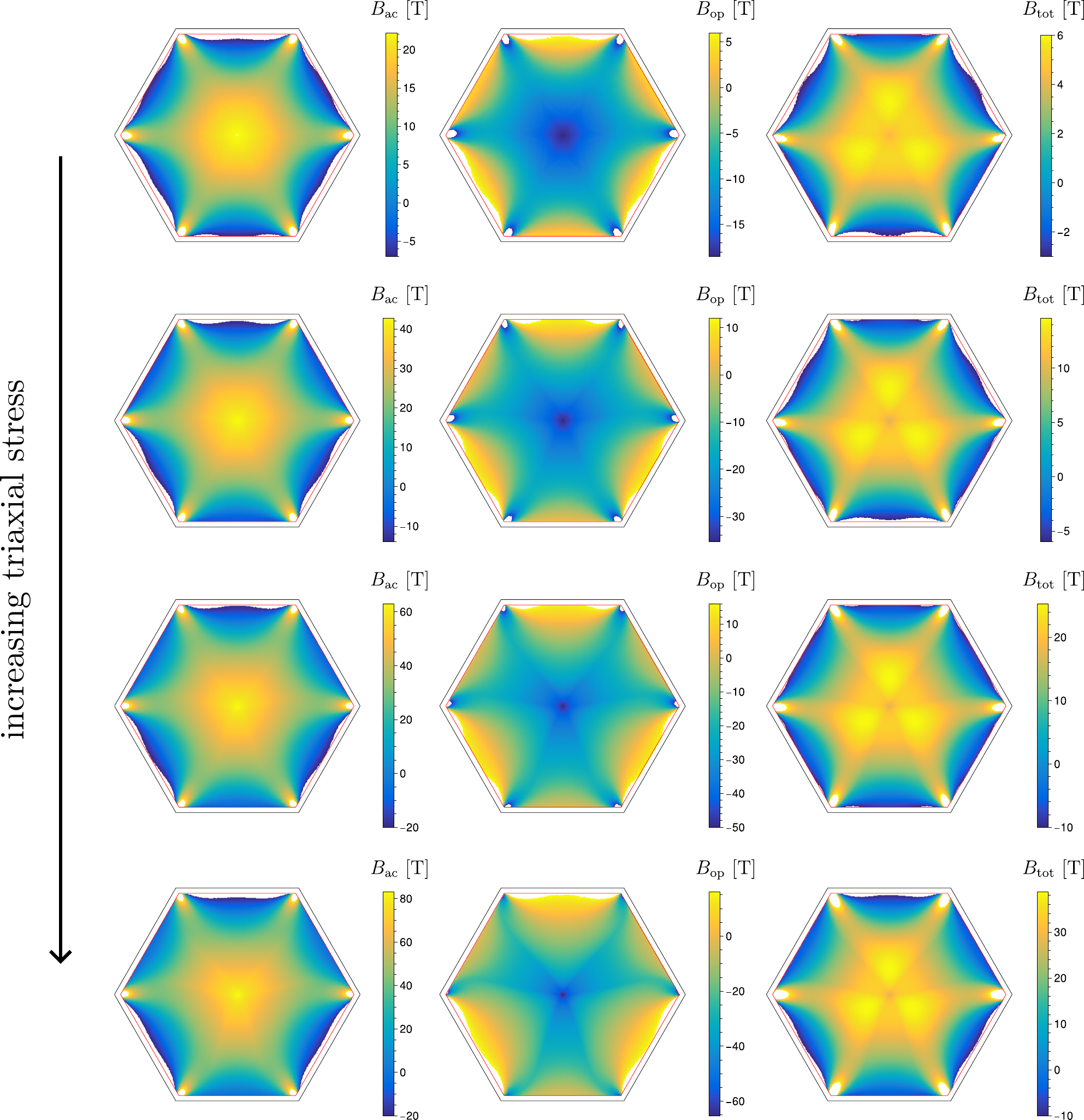}
    \caption{Acoustic, optical, and total PMF for a hexagonal graphene flake with zigzag edges (black hexagon) under increasing triaxial stress, where the horizontal direction is the zigzag direction. Only sites in the smaller red hexagon are included in the calculation to avoid boundary effects.}
    \label{fig:hexagon}
\end{figure*}

In this section, we calculate the different contributions to the PMF for a finite hexagonal graphene flake subjected to in-plane triaxial stress \cite{neek-amal_electronic_2013}. We achieve this by first extracting the displacements field $\bm u_\sigma$ ($\sigma = A, B$) and performing a linear interpolation on a triangular grid. Then we calculate the acoustic and optical fields $\bm u$ and $\bm v$, respectively. With these fields we construct two new structures that only have acoustic or only have optical displacements. We then calculate the effective vector potential using the definition from Eq.\ \eqref{eq:Adef} of the main text in terms of the $\delta t_n$. Here we used the hopping amplitude given in Eq.\ \eqref{eq:hopping}  of the main text. After another linear interpolation we then calculate the pseudomagnetic field for the original structure, the structure with only acoustic displacements, and the structure with only optical displacements. The results for four increasing values of the triaxial stress for an hexagonal graphene flake with zigzag edges are shown in Fig.\ \ref{fig:hexagon}. At the center of the flake, we find a reduction factor of about $4$ which matches well with Fig.\ 4(f) of Ref.\ \cite{neek-amal_electronic_2013}.

\end{document}